\documentclass[a4paper,11pt]{article}

\usepackage{jheppub}
\usepackage[english]{babel}
\usepackage[dvipsnames]{xcolor}
\usepackage[T1]{fontenc}
\usepackage{cite}
\usepackage{relsize,scalerel}
\usepackage[export]{adjustbox}
\usepackage{amsmath, amsfonts, amssymb,wasysym}
\usepackage{slashed}
\usepackage{simpler-wick}
\usepackage{graphicx,tikz,pgf}
\usetikzlibrary{cd,shapes,decorations.pathmorphing,decorations.markings,decorations.pathreplacing}
\usepackage{ifthen,xparse}
\usepackage{physics}
\usepackage{soul}
\usepackage{braket}
\usepackage{bm}
\usepackage{siunitx}
\usepackage{stackrel} 
\usepackage{subcaption}
\usepackage[mathlines]{lineno}
\usepackage{mathtools}
\allowdisplaybreaks

\begin{document}
\title{Inclusive hadroproduction of $\chi_{c1}(3872)$, $X_b$ and pentaquarks}

\author[a,b,c]{Nora Brambilla,}
\author[d]{Mathias Butenschoen,}
\author[a]{Simon Hibler,}
\author[a]{Abhishek Mohapatra,}
\author[a]{Antonio Vairo,}
\author[e]{Xiangpeng Wang}
\affiliation[a]{Technical University of Munich, TUM School of Natural Sciences, Physics Department,\\ James-Franck-Str.~1, 85748 Garching, Germany}
\affiliation[b]{Technical University of Munich, Institute for Advanced Study, \\ 
Lichtenbergstrasse 2 a, 85748 Garching, Germany}
\affiliation[c]{Technical University of Munich, Munich Data Science Institute, \\ 
Walther-von-Dyck-Strasse 10, 85748 Garching, Germany}
\affiliation[d]{II. Institut f\"ur Theoretische Physik, Universit\"at Hamburg, \\
Luruper Chaussee 149, 22761 Hamburg, Germany}
\affiliation[e]{Institute of Particle Physics and Key Laboratory of Quark and Lepton Physics (MOE), \\
Central China Normal University, Wuhan, Hubei 430079, China}

\emailAdd{nora.brambilla@tum.de}
\emailAdd{mathias.butenschoen@desy.de}
\emailAdd{simon.hibler@tum.de}
\emailAdd{abhishek.mohapatra@tum.de}
\emailAdd{antonio.vairo@tum.de}
\emailAdd{xpwang@ccnu.edu.cn}

\abstract{We use the Born--Oppenheimer effective field theory factorization to compute the inclusive production cross sections of the $\chi_{c1}(3872)$ and its partner in the bottomonium sector.
In the same framework, we compute the production cross sections of the pentaquark states $P_{c\bar{c}}(4312)^+$, $P_{c\bar{c}}(4457)^+$, $P_{c\bar{c}}(4380)^+$ and $P_{c\bar{c}}(4440)^+$ within two possible scenarios for the Born--Oppenheimer potentials.
Also for pentaquarks, we extend the results to the bottomonium sector. 
All our results are genuine predictions that do not involve fits to prompt  hadroproduction data.
}

\maketitle
\newpage

\section{Introduction}
Over the past two decades, following the discovery of the $\chi_{c1}(3872)$ state by the Belle experiment~\cite{Belle:2003nnu}, a substantial number of new states have been observed across both the charmonium and bottomonium spectrum~\cite{Brambilla:2019esw}. 
The properties of these states, collectively known as XYZs, cannot be explained by ordinary quarkonia, suggesting  more complicated structures.
These include tetraquarks and pentaquarks.

States made by two heavy (anti)quarks are most conveniently described by 
nonrelativistic effective field theories of QCD~\cite{Brambilla:2004jw}.
They exploit the hierarchy of energy scales typical of nonrelativistic bound states.
These are $m_Q$, the mass of a heavy quark of flavor~$Q$, $m_Q v$, and $m_Q v^2$, $v \ll 1$ being the heavy quark velocity in the quarkonium rest frame. 
Nonrelativistic effective field theories express observables as expansions in $v$.
In this paper, we compute at leading order in $v$ the inclusive cross sections of the $\chi_{c1}(3872)$, the observed non-strange pentaquarks with hidden charm, and their partners in the bottomonium sector by making use of the factorization formulas provided by {\it nonrelativistic QCD} (NRQCD), {\it potential nonrelativistic QCD} (pNRQCD) and the {\it Born--Oppenheimer effective field theory} (BOEFT).

Nonrelativistic QCD is the nonrelativistic effective field theory that follows from QCD by integrating out modes of energy or momentum of order $m_Q$.
Inclusive quarkonium production cross sections are factorized into short-distance coefficients, encoding the contributions from the scale $m_Q$, and long-distance matrix elements, encoding the contributions from the low-energy modes~\cite{Bodwin:1994jh}. 
Short-distance coefficients are computed in perturbation theory, order by order in the strong coupling~$\alpha_s$. 
Long-distance matrix elements are nonperturbative quantities that are determined by fitting cross sections and other observables to data.

Potential nonrelativistic QCD is the effective field theory that follows from NRQCD by integrating out gluons with energy or momentum of order $m_Qv$.
Strongly coupled pNRQCD is the version of pNRQCD that is suitable to describe the inclusive production of charmonium and excited bottomonium states whose binding energy, which is of order $m_Q v^2$, is smaller than the hadronic scale $\Lambda_{\rm QCD}$~\cite{Brambilla:1999xf,Brambilla:2000gk,Pineda:2000sz,Brambilla:2001xy,Brambilla:2002nu}. 
In strongly coupled pNRQCD, the long-distance matrix elements of NRQCD are factorized into a quarkonium-dependent part that is the square of the quarkonium wave function or its derivatives at the origin and some matrix elements or correlators that do not depend on the quarkonium state~\cite{Brambilla:2020ojz,Brambilla:2021abf,Brambilla:2022rjd,Brambilla:2022ayc}.
Because of the universal nature of the latter, long-distance matrix elements of different quarkonia can be related to one another, thereby reducing the overall number of nonperturbative unknowns significantly. 
The extension of pNRQCD to describe the XYZ states is called Born--Oppenheimer effective field theory~\cite{Berwein:2015vca,Oncala:2017hop,Brambilla:2017uyf,Soto:2020xpm,Berwein:2024ztx,Braaten:2024tbm}. 

The $\chi_{c1}(3872)$ is most likely a tetraquark with hidden charm~\cite{Brambilla:2019esw}.
In~\cite{Berwein:2024ztx}, it has been shown that the charm-anticharm pair in the $\chi_{c1}(3872)$ is at short distance in a color octet configuration.
Since production is a process that happens at a short distance of order $1/m_Q$, 
the production of the $\chi_{c1}(3872)$ involves, at leading order in $v$, the long-distance matrix element of an operator projecting on color octet charm-anticharm states.
In analogy to pNRQCD for the quarkonium case, the BOEFT for the $\chi_{c1}(3872)$ case allows to factorize the long-distance matrix element into a wave function part and some universal matrix element~\cite{Lai:2025tpw}.
The wave function can be determined by solving the Schr\"odinger equations that follow from the leading order 
equations of motion of the BOEFT for the  $\chi_{c1}(3872)$ state~\cite{Berwein:2024ztx,Brambilla:2024imu}.
The universal matrix element may be fitted to $B$ hadron decay data and used to make genuine predictions for the hadroproduction of both the $\chi_{c1}(3872)$ and the state analogous to the $\chi_{c1}(3872)$ in the bottomonium sector, called $X_b$.
The detailed derivation of the $\chi_{c1}(3872)$ inclusive production cross section at leading order in $v$ in the BOEFT, 
and the consequent phenomenological analyses with LHC kinematics in the $\chi_{c1}(3872)$ and $X_b$ cases are the subject of the first part of the paper.
In the second part of the paper, we extend the study to the non-strange pentaquarks with hidden charm recently discovered at the LHCb experiment in the charmonium spectrum, and to their yet-to-be-discovered partners in the bottomonium spectrum.

In more detail, the paper is organized in the following way.
In section~\ref{sec:QQbar}, we briefly review the NRQCD and pNRQCD factorization of quarkonium inclusive production on the example of $P$-wave quarkonium.
By extending pNRQCD to BOEFT, we derive factorization formulas for the production of the $\chi_{c1}(3872)$ in section~\ref{sec:QQbarqqbar} and for the production of the pentaquark states 
$P_{c\bar{c}}(4312)^+$, $P_{c\bar{c}}(4457)^+$, $P_{c\bar{c}}(4380)^+$ and $P_{c\bar{c}}(4440)^+$ in section~\ref{sec:QQbarqqq}. 
Moreover, in section~\ref{sec:QQbarqqbar}, we provide predictions for the inclusive production cross sections of the $\chi_{c1}(3872)$ and its partner in the bottomonium sector, $X_b$. 
Similarly, in section~\ref{sec:QQbarqqq}, we predict inclusive production cross sections for pentaquarks with hidden charm and hidden bottom. 
Finally, in section~\ref{sec:conclusions}, we summarize our findings and draw some conclusions.

\section{Inclusive production of $P$-wave quarkonium} 
\label{sec:QQbar}
Before looking at the production of the tetraquark state $\chi_{c1}(3872)$, we briefly summarize the case of $P$-wave quarkonium production.
The starting point is the nonrelativistic QCD factorization formula for the inclusive quarkonium production cross section~\cite{Bodwin:1994jh}
\begin{align}
    \sigma_{{\cal Q}}=\sum_N \sigma_{Q\bar{Q}(N)} \bra{\Omega}\mathcal{O}^{\cal Q}(N)\ket{\Omega},
    \label{eq:NRQCD_Factorized}
\end{align}
where $\cal Q$ stands for the produced quarkonium state, $\sigma_{Q\bar{Q}(N)}$ are {\it short-distance coefficients} encoding the production cross sections of heavy quark-antiquark pairs, $Q\bar{Q}$, in a state $N={}^{2S+1}L_J^{[1,8]}$, with $[1], [8]$ labeling color-singlet and color-octet configurations, respectively, and $\ket{\Omega}$ is the QCD vacuum state. 
For the production of $P$-wave quarkonia, $\chi_{QJ}$, the two relevant {\it long-distance matrix elements} (LDMEs) at leading order in $v$ are  
\begin{align}
   \bra{\Omega} {\cal O}^{\chi_{Q0}} ({}^3P_0^{[1]}) \ket{\Omega} &= \frac{1}{3} \bra{\Omega} \chi^\dagger (- \tfrac{i}{2} \overleftrightarrow{\bm{D}} \cdot \bm{\sigma}) \psi \, {\cal P}_{\chi_{Q0}} \, \psi^\dagger (- \tfrac{i}{2} \overleftrightarrow{\bm{D}} \cdot \bm{\sigma}) \chi \ket{\Omega}, 
   \label{eq:OperatorP}\\
   \bra{\Omega}\mathcal{O}^{\chi_{Q0}}({}^3S_1^{[8]})\ket{\Omega}&=\bra{\Omega}\chi^\dagger \sigma^k T^A\psi\Phi_\ell^{\dagger AB} \, {\cal P}_{\chi_{Q0}} \,\Phi_\ell^{BC}\psi^\dagger\sigma^kT^C\chi\ket{\Omega}, 
   \label{eq:OperatorS} 
\end{align}
where the Pauli field $\psi^\dagger$ creates a heavy quark, the Pauli field $\chi$ creates a heavy antiquark,
$D^i$ are the spatial components of the gauge covariant derivative, 
$T^A$ are color matrices, $\sigma^k$ are the Pauli matrices, 
${\cal P}_{\chi_{Q0}}$ is an operator projecting onto states that contain a $\chi_{Q0}$ quarkonium at rest 
and $\Phi_\ell$ is an adjoint Wilson line along the direction $\ell$, necessary to ensure the gauge invariance of the octet matrix element~\cite{Nayak:2005rw,Nayak:2005rt}.
The other LDMEs for $P$-waves belonging to the same spin multiplet are related by heavy quark spin symmetry.
In the NRQCD power counting, which is based on the expansion in the small heavy quark velocity $v$ in the quarkonium rest frame, 
the two matrix elements are both of order $v^2$ (counting of order $v^0$ the contribution of unsuppressed dimension six  four-fermion operators).
The order $v^2$ suppression of the color-singlet LDME in \eqref{eq:OperatorP} originates from the two derivatives, while the $v^2$ suppression of the color-octet LDME in \eqref{eq:OperatorS} comes from the color-octet components of the quarkonium states in the projector ${\cal P}_{\chi_{QJ}}$.
At leading order in $v$, the quarkonium $P$-wave production cross section \eqref{eq:NRQCD_Factorized} becomes therefore 
\begin{equation}
\label{eq:NRQCDfacQuarkonium}
\sigma_{\chi_{QJ}} =  \sigma_{Q \bar Q({}^3P_J^{[1]})} \bra{\Omega} {\cal O}^{\chi_{QJ}} ({}^3P_J^{[1]}) \ket{\Omega} 
+ \sigma_{Q \bar Q({}^3S_1^{[8]})}\bra{\Omega} {\cal O}^{\chi_{QJ}} ({}^3S_1^{[8]}) \ket{\Omega}.
\end{equation}
If the polarization of the produced quarkonium states is summed in the LDMEs, then the cross section is an {\it unpolarized cross section}.
Here and in the following, we assume this to be the case, which, owing to the spin symmetry, implies that $   \bra{\Omega} {\cal O}^{\chi_{QJ}} ({}^3P_J^{[1]}) \ket{\Omega} = (2J+1)\bra{\Omega} {\cal O}^{\chi_{Q0}} ({}^3P_0^{[1]}) \ket{\Omega}$ and $ \bra{\Omega}\mathcal{O}^{\chi_{QJ}}({}^3S_1^{[8]})\ket{\Omega} = (2J+1)\bra{\Omega}\mathcal{O}^{\chi_{Q0}}({}^3S_1^{[8]})\ket{\Omega}$.

In pNRQCD, the $P$-wave quarkonium LDMEs can be written in terms of the derivative at the origin of the radial part of the quarkonium wave function, $\phi_{\chi_{QJ}}$, and a universal correlator.
The result in the case of the NRQCD LDMEs listed above is, up to corrections of order $v^2$ and $1/N_c^2$ in the case of the octet LDME,~\cite{Brambilla:2020ojz,Brambilla:2021abf}
\begin{align}
\bra{\Omega} {\cal O}^{\chi_{Q0}} ({}^3P_0^{[1]}) \ket{\Omega}  &= \frac{3 N_c}{2 \pi} | \phi^{\prime}_{\chi_{Q0}}(0)|^2 ,\label{eq:O1pNRQCD}\\
\bra{\Omega} {\cal O}^{\chi_{Q0}} ({}^3S_1^{[8]}) \ket{\Omega}  &= \frac{3 N_c}{2 \pi} | \phi^{\prime}_{\chi_{Q0}}(0)|^2 \, \frac{{\cal E}(\Lambda)}{9 N_c m_Q^2},
\label{eq:O8pNRQCD}
\end{align}
where $N_c=3$ is the number of colors, and $\mathcal{E}(\Lambda)$ is given by 
\begin{align}
    {\cal E}(\Lambda) = \frac{3}{N_c} \int_0^\infty dt\, t \int_0^\infty dt'\, t' \, \bra{\Omega} \Phi_\ell^{\dag ab} \Phi_0^{\dag da} (0,t) g E^{d,i} (t) g E^{e,i} (t') \Phi^{ec}_0 (0,t') \Phi_\ell^{bc} \ket{\Omega}.
\end{align}
The result \eqref{eq:O1pNRQCD} for the color singlet LDME can be understood as a consequence of approximating the projector ${\cal P}_{\chi_{Q0}}$ with a projector on the $\chi_{c0}$ state plus the vacuum alone, 
an approximation that goes under the name of {\it vacuum saturation approximation} in NRQCD 
and holds up to order $v^2$ corrections~\cite{Bodwin:1994jh}.
The result \eqref{eq:O8pNRQCD} is a genuine result of pNRQCD.
The correlator ${\cal E}(\Lambda)$ depends on the renormalization scale $\Lambda$.
It has been extracted at $\Lambda = 1.5$~GeV in~\cite{Brambilla:2021abf}; 
evolving it to $m_c = 1.4 \, \mathrm{GeV}$ (see footnote~\ref{footnotemass}) at leading logarithmic accuracy results in $\mathcal{E}(\Lambda = 1.4 \, \mathrm{GeV}) = 2.5 \pm 1.7$. 
Similarly, we can calculate the LDMEs for the production of exotic hadrons.

\section{Inclusive production of $\chi_{c1}(3872)$ and $X_b$}
\label{sec:QQbarqqbar}
In this section, we investigate the production of the state $\chi_{c1}(3872)$.
The $\chi_{c1}(3872)$ is a well established state found in the charmonium spectrum~\cite{Belle:2003nnu,LHCb:2013kgk,LHCb:2015jfc,LHCb:2020fvo} 
with a likely large hidden charm tetraquark component, for a review, see~\cite{Brambilla:2019esw}. 
Its quantum numbers are $J^{PC}=1^{++}$ with isospin $I=0$. 

Tetraquarks and other non-conventional quarkonium states like hybrids, pentaquarks, or doubly heavy baryons can be described in a systematic fashion within the Born--Oppenhei\-mer effective field theory~\cite{Berwein:2015vca,Oncala:2017hop,Brambilla:2017uyf,Soto:2020xpm,Berwein:2024ztx,Braaten:2024tbm}.
The Born--Oppenheimer effective field theory is a nonrelativistic effective field theory of QCD that extends  
potential NRQCD to describe, besides conventional quarkonia, also non-conventional states made of a heavy quark-(anti)quark pair. 
The light degrees of freedom of conventional and non-conventional quarkonia are classified according to the symmetry group of diatomic molecules.
They are labeled $\Lambda^\sigma_\eta$, 
where $\Lambda = |\bm{r}\cdot\bm{k}| = 0\,(\equiv \Sigma),\,1/2,\,1\,(\equiv \Pi), \,3/2,\,2\,(\equiv\Delta),\, 5/2,\, ...$ is the projection of the angular momentum $\bm{k}$ of the light degrees of freedom on the distance $\bm{r}$ between the heavy quark and (anti)quark, 
$\eta$ is the parity P or CP (for states symmetric under charge conjugation C) eigenvalue of the light degrees of freedom ($g\equiv 1$ and $u\equiv -1$), 
and $\sigma$ is their reflection eigenvalue (only for $\Sigma$ states).
The numbers $\Lambda$, $\sigma$ and $\eta$ are called {\it Born--Oppenheimer (BO) quantum numbers}.
Excited states with the same BO quantum numbers are labeled by primes, e.g. $\Sigma_g$, $\Sigma_g^\prime$,~...~.
The BOEFT realizes at first order in the nonrelativistic expansion the Born--Oppenheimer approximation.
In the Born--Oppenheimer approximation, the heavy quarks move adiabatically in the presence of the light degrees of freedom,
whose effect is encoded in a suitable set of potentials that depend on the distance $\bm{r}$. 
The equations of motion are simple or coupled Schr\"odinger equations.

The Born--Oppenheimer quantum numbers for the light degrees of freedom of the $\chi_{c1}(3872)$ are $\Sigma_g^{+\prime}$ and $\Pi_g$, where $g$ stands here for even under CP.
The states $\Sigma_g^{+\prime}$ and $\Pi_g$ mix at short distance where they become degenerate with an adjoint meson of quantum numbers $k^{PC}=1^{--}$ due to the restoration of spherical symmetry~\cite{Berwein:2024ztx}.
Moreover, at large distance, the light degrees of freedom with quantum numbers $\Sigma_g^{+\prime}$ mix with the quarkonium light degrees of freedom that have BO quantum numbers $\Sigma_g^+$, resulting in avoided level crossing.
These mixing patterns lead to the following coupled (radial) Schr\"odinger equations describing the $\chi_{c1}(3872)$ and the $\chi_c(1P)$ charmonium state~\cite{Berwein:2024ztx,Brambilla:2024imu}:
\begin{align}
&\left[
-\frac{1}{m_cr^2}\,\partial_rr^2\partial_r+\frac{1}{m_cr^2}
{\begin{pmatrix}
2 & 0 & 0\\
0                 & 4        & -2\sqrt{2} \\
0                 & -2\sqrt{2} & 2
\end{pmatrix}}\right.
\nonumber\\
&\hspace{3.5 cm}\left.
+\begin{pmatrix} V_{\Sigma_{g}^{+}}(r) &  V_{\Sigma_{g}^{+}-\Sigma_{g}^{+\prime}}(r) & 0 \\
    V_{\Sigma_{g}^{+}-\Sigma_{g}^{+\prime}}(r) & V_{\Sigma_{g}^{+\prime}}(r) & 0\\
      0 & 0 & V_{\Pi_g}(r)\end{pmatrix}
      \right]
  \hspace{-4pt}\begin{pmatrix} \phi_{\Sigma^+_g}    \\ \phi_{\Sigma^{+\prime}_g} \\ \phi_{\Pi_g }\end{pmatrix}= 
  E \begin{pmatrix} \phi_{\Sigma^+_g} \\ \phi_{\Sigma^{+\prime}_g} \\ \phi_{\Pi_g}\end{pmatrix},
\label{coupledI0}
\end{align}
where $m_c$ is the charm mass,\footnote{
The scheme for the mass entering \eqref{coupledI0} depends on the scheme adopted for the potentials.
}  
$\bm{r}$ is the relative distance of the heavy quark-antiquark pair, 
$V_{\Sigma_{g}^{+}}(r)$, $V_{\Sigma_{g}^{+\prime}}(r)$ and $V_{\Pi_g}(r)$ are the BO potentials, and 
$V_{\Sigma_{g}^{+}-\Sigma_{g}^{+\prime}}(r)$ is the $\Sigma_{g}^{+}$-$\Sigma_{g}^{+\prime}$ mixing potential.
These potentials are known or partially known from lattice determinations and symmetry constraints, 
so that eq.~\eqref{coupledI0} can be (and has been) solved to give the radial parts 
of the wave functions, $\phi_{\Sigma^+_g}$, $\phi_{\Sigma^{+\prime}_g}$ and $\phi_{\Pi_g}$, 
contributing to the eigenstate with binding energy $E$~\cite{Berwein:2024ztx,Brambilla:2024imu}. 
The radial wave function $\phi_{\Sigma^+_g}$ provides the quarkonium component of the state, while the radial wave functions $\phi_{\Sigma^{+\prime}_g}$ and $\phi_{\Pi_g}$ provide the two tetraquark components.

For the $\chi_{c1}(3872)$, the quarkonium component, $\displaystyle \int_0^\infty dr\,r^2 \,|\phi_{\Sigma^+_g}(r)|^2$, is small~\cite{Brambilla:2024imu}, 
which makes the $\chi_{c1}(3872)$ mostly a tetraquark state. 
The light degrees of freedom of the tetraquark state become a $1^{--}$ adjoint meson at short distance, which implies that the $Q\bar{Q}$ pair in the $\chi_{c1}(3872)$ is produced in a color octet configuration.
The production mechanism of the $\chi_{c1}(3872)$ is therefore different from the production mechanism of the $P$-wave quarkonium reviewed in section~\ref{sec:QQbar}.\footnote{
Production of the $\chi_{c1}(3872)$ through its $P$-wave quarkonium component has been considered in~\cite{Butenschoen:2013pxa,Meng:2013gga,Butenschoen:2019npa}.}
For $\chi_{c1}(3872)$ production, the octet matrix element gives the dominant contribution, whereas the singlet matrix element is not only suppressed by $v^2$ through the two covariant derivatives, 
but also by the small quarkonium component in the $\chi_{c1}(3872)$ state.
At leading order in $v$, the inclusive $\chi_{c1}(3872)$ production cross section reads in terms of the NRQCD factorization formula \eqref{eq:NRQCD_Factorized}
\begin{equation}
\sigma_{\chi_{c1}(3872)} = \sigma_{Q \bar Q({}^3S_1^{[8]})}\bra{\Omega} {\cal O}^{\chi_{c1}(3872)} ({}^3S_1^{[8]}) \ket{\Omega}, 
\label{eq:NRQCDfacX3872}
\end{equation}
with the octet LDME 
\begin{equation}
 \bra{\Omega}\mathcal{O}^{\chi_{c1}(3872)}({}^3S_1^{[8]})\ket{\Omega} = 
 \bra{\Omega}\chi^\dagger \sigma^k T^A\psi\Phi_\ell^{\dagger AB} \, {\cal P}_{\chi_{c1}(3872)} \,\Phi_\ell^{BC}\psi^\dagger\sigma^k T^C\chi\ket{\Omega}.
\label{eq:OperatorX3872} 
\end{equation}

The octet LDME depends on the projection operator ${\cal P}_{\chi_{c1}(3872)}$.
In order to write the projection operator we need, first, to write the state that describes the $\chi_{c1}(3872)$. 
In a generic reference frame so that $\bm{P}$ is the $\chi_{c1}(3872)$ center of mass momentum, 
the state is given by~\cite{Berwein:2024ztx,Brambilla:2024imu}
\begin{align}
&     \ket{\chi_{c1}(3872);\bm{P}} = 
     \sum_{m_l,\,m_S}\,C_{J=1,\,m_J;\,l=1,\,S=1}^{\,m_l,\,m_S}
     \int d^3R\,d^3r\,e^{i\bm{P}\cdot \bm{R}}  \, \bigg\{ \nonumber\\
     &\hspace{15mm}  \psi_{\alpha,i}^{\dagger}(\bm{R} + \bm{r}/2)\chi_{\beta, j}(\bm{R} - \bm{r}/2) \,(\Psi_{\Sigma^+_g}^{\lambda=0})^{l=1,m_l;S=1,m_S}_{\alpha\beta}(\bm{r})\ket{0;0^{++},\Sigma_g^+(\lambda=0); \bm{r}; i,j} \nonumber\\
     &\hspace{10mm}+ \psi_{\alpha,i}^{\dagger}(\bm{R} + \bm{r}/2)\chi_{\beta, j}(\bm{R} - \bm{r}/2)
     \bigg[(\Psi_{\Sigma_g^{+\prime}}^{\lambda=0})_{\alpha\beta}^{l=1,m_l;S=1,m_S}(\bm{r})\ket{0;1^{--}, \Sigma_g^{+\prime}(\lambda=0); \bm{r}; i,j} \nonumber\\
     &\hspace{28mm} +\frac{1}{\sqrt{2}}\bigg((\Psi_{\Pi_g}^{\lambda=1})_{\alpha\beta}^{l=1,m_l;S=1,m_S}(\bm{r})\ket{0;1^{--}, \Pi_g(\lambda=1); \bm{r}; i,j}\nonumber\\
     &\hspace{38mm} +(\Psi_{\Pi_g}^{\lambda=-1})_{\alpha\beta}^{l=1,m_l;S=1,m_S}(\bm{r})\ket{0;1^{--}, \Pi_g(\lambda=-1); \bm{r}; i,j}\bigg)\bigg] \bigg\}
\nonumber\\
& \hspace{10mm} \approx 
\sum_{m_l,\,m_S}\,C_{J=1,\,m_J;\,l=1,\,S=1}^{\,m_l,\,m_S}
     \int d^3R\,d^3r\,e^{i\bm{P}\cdot \bm{R}}  \, \bigg\{ \nonumber\\
     & \hspace{15mm} \psi_{\alpha,i}^{\dagger}(\bm{R} + \bm{r}/2)\chi_{\beta, j}(\bm{R} - \bm{r}/2)
     \bigg[(\Psi_{\Sigma_g^{+\prime}}^{\lambda=0})_{\alpha\beta}^{l=1,m_l;S=1,m_S}(\bm{r})\ket{0;1^{--}, \Sigma_g^{+\prime}(\lambda=0); \bm{r}; i,j} \nonumber\\
     &\hspace{28mm} +\frac{1}{\sqrt{2}}\bigg((\Psi_{\Pi_g}^{\lambda=1})_{\alpha\beta}^{l=1,m_l;S=1,m_S}(\bm{r})\ket{0;1^{--}, \Pi_g(\lambda=1); \bm{r}; i,j}\nonumber\\
     &\hspace{38mm} +(\Psi_{\Pi_g}^{\lambda=-1})_{\alpha\beta}^{l=1,m_l;S=1,m_S}(\bm{r})\ket{0;1^{--}, \Pi_g(\lambda=-1); \bm{r}; i,j}\bigg)\bigg] \bigg\}.     
     \label{eq:X3872state}  
\end{align}
The total angular momentum of the state is $\bm{J}=\bm{L}+\bm{S}$, with $\bm{L}=\bm{L}_{Q\bar{Q}}+\bm{K}$ the 
sum of the orbital angular momentum of the $Q\bar{Q}$ pair, ${\bm L}_{Q\bar{Q}}$, and the angular momentum of the light degrees of freedom, ${\bm K}$, and ${\bm S}={\bm S_1}+{\bm S_2}$ the total spin of the $Q\bar{Q}$ pair. 
The eigenvalues of $\bm{J}^2$ and $J_3$ are $J(J+1)$ and $m_J$, 
the eigenvalues of $\bm{L}^2$ and $L_3$ are $l(l+1)$ and $m_l$, 
and the eigenvalues of $\bm{S}^2$ and $S_3$ are $S(S+1)$ and $m_S$, respectively. 
The factors $C_{J,\,m_J;\,l,\,S}^{\,m_l,\,m_S}$ are the Clebsch--Gordan coefficients that allow to construct out of the eigenstates of $L_3$ and $S_3$ eigenstates of $J$ and $J_3$. 
Some quantum numbers are fixed for the physical state $\chi_{c1}(3872)$.
These are $J=1$, $l=1$, and $S=1$~\cite{Brambilla:2024imu}; moreover, parity and charge conjugation are positive. 
The three eigenvalues of $J_3$ provide the three polarizations of the state.
The indices $\alpha$ and $\beta$, which may assume the values 1 or 2,  
are the spinor indices, and the indices $i,j$, which may assume the values from 1 to 3, are the color indices. 
The wave functions $\Psi_{\Lambda}^{\lambda}(\bm{r})$, with $\Lambda = \Sigma_g^+$, $\Sigma_g^{+\prime}$ or $\Pi_g$ are related to the radial parts appearing in the Schr\"odinger equations \eqref{coupledI0} by 
\begin{equation}
    (\Psi_{\Lambda}^{\lambda})^{l=1,m_l;S=1,m_S}_{\alpha\beta}(\bm{r}) = 
    v_{l=1,m_l}^{\lambda}\,\phi_{\Lambda}(r) \,\frac{{\bm e}_{m_S}\cdot \bm{\sigma}_{\alpha\beta}}{\sqrt{2}},
\end{equation}
where $v_{l=1,m_l}^{\lambda}$ are the angular wave functions~\cite{Berwein:2015vca} and $\bm{e}_{m_S}$ are three unit vectors identifying the three polarizations of a spin $S=1$ state.

The $\Psi_{\Sigma_g^+}^{\lambda=0}(\bm{r})$ term in~\eqref{eq:X3872state} gives the quarkonium component, whereas the remaining terms give the tetraquark components. 
For the $\Pi_g$ tetraquark component, we have selected the positive parity combination. 
In the last (approximate) equality, we have neglected the quarkonium component, which is small compared to the tetraquark one in the $\chi_{c1}(3872)$ and, moreover, suppressed in the production cross section, as we argued above.
The Pauli field $\psi^\dagger(\bm{R}+\bm{r}/2)$ creates a heavy quark at $\bm{R}+\bm{r}/2$ and the Pauli field $\chi(\bm{R}-\bm{r}/2)$ creates a heavy antiquark at $\bm{R}-\bm{r}/2$; 
$\bm{R}$ is the center of mass coordinate and $\bm{r}$ the relative coordinate. 
The state $\ket{0;0^{++},\Sigma_g^+(\lambda=0);\bm{r}; i,j}$, and the two states  
$\ket{0;1^{--}, \Sigma_g^{+\prime}(\lambda=0); \bm{r}; i,j}$ and $\ket{0;1^{--}, \Pi_g(\lambda=\pm 1); \bm{r}; i,j}$ 
encode the light degrees of freedom of the quarkonium component and tetraquark components  
of the state $\ket{\chi_{c1}(3872)}$, respectively; these states do not contain heavy quarks.
They are labeled in the following way: 
the first 0 means that we do not add to the $\chi_{c1}(3872)$ extra final state particles $X$,
the second entry provides the $k^{PC}$ quantum numbers of the light degrees of freedom forming the $\chi_{c1}(3872)$ at short distance, 
the third entry provides the BO quantum numbers, where we also specify in parentheses the polarization $\lambda$, 
the fourth entry makes explicit the $\bm{r}$ dependence,\footnote{
For simplicity, we have dropped the dependence on the center of mass coordinate, $\bm{R}$. 
This may eventually be set to $\bm{0}$ when computing the LDMEs because of translational invariance.}
and the last two entries show the color indices.
The states $\ket{0;0^{++},\Sigma_g^+(\lambda=0);\bm{r}; i,j}$, $\ket{0;1^{--}, \Sigma_g^{+\prime}(\lambda=0); \bm{r}; i,j}$ and $\ket{0;1^{--}, \Pi_g(\lambda=\pm 1); \bm{r}; i,j}$ are orthonormal at the same point.
As a consequence, the state $ \ket{\chi_{c1}(3872);\bm{P}}$ is normalized in the usual way of nonrelativistic states:
$\bra{\chi_{c1}(3872);\bm{P}} \chi_{c1}(3872);\bm{P}'\rangle = (2\pi)^3\delta^3(\bm{P}-\bm{P}')$.
The quantum number $m_J$ labels the polarizations of the state.
Computing the unpolarized projection operator and the unpolarized LDME requires summing over the three polarizations $m_J$.
The radial parts of the wave functions, $\phi_{\Lambda}(\bm r)$, are the solutions of the coupled Schr\"odinger equations~\eqref{coupledI0}.
The explicit form of the states  $\ket{0;0^{++},\Sigma_g^+(\lambda=0);\bm{r}; i,j}$, $\ket{0;1^{--}, \Sigma_g^{+\prime}(\lambda=0); \bm{r}; i,j}$ and $\ket{0;1^{--}, \Pi_g(\lambda=\pm 1); \bm{r}; i,j}$ is unknown in general, but it is known at short distance.
In particular, we have that at leading order in the velocity expansion~\cite{Brambilla:2020ojz,Brambilla:2021abf}
\begin{equation}
\ket{0;0^{++},\Sigma_g^+(\lambda=0);\bm{0};i,j} = \frac{\delta_{ij}}{\sqrt{N_c}} \ket{\Omega},
\label{eq:shortsinglet}
\end{equation}
and~\cite{Berwein:2024ztx}
\begin{align}
\ket{0;1^{--},\Sigma_g^{+\prime}(\lambda=0);\bm{0};i,j} &= \nonumber\\
& \hspace{-13mm} \left(\frac{2}{ \left|\bar{q}(\bm{0})  \bm{P}_{10} \cdot \bm{\gamma} \, T^B q(\bm{0})\ket{\Omega}\right|^2}\right)^{1/2} T^A_{ij} \bar{q}(\bm{R}) \bm{P}_{10} \cdot \bm{\gamma} \, T^A q(\bm{R}) \ket{\Omega},
\label{eq:shorttetra1}\\
\ket{0;1^{--},\Pi_g(\lambda=\pm 1);\bm{0};i,j} &= \nonumber\\
& \hspace{-13mm} \left(\frac{2}{\left|\bar{q}(\bm{0})  \bm{P}_{1\pm1} \cdot \bm{\gamma} \, T^B q(\bm{0})\ket{\Omega}\right|^2}\right)^{1/2} T^A_{ij} \bar{q}(\bm{R}) \bm{P}_{1\pm 1} \cdot \bm{\gamma} \, T^A q(\bm{R}) \ket{\Omega},
\label{eq:shorttetra2}
\end{align}
where $q$ are the light quark fields in the tetraquark, $\bm{P}_{10} = \hat{\bm{r}}$ and $\bm{P}_{1\pm1} = \mp (\hat{\bm{\theta}} \pm i \hat{\bm{\varphi}})/\sqrt{2}$ for the two degenerate $\Pi_g$ states.\footnote{
$\bm{P}_{10}$ is the unit vector $\bm{r}/r$ and $\bm{P}_{1\pm1}$ are unit vectors orthogonal to $\bm{r}$ expressed in terms of spherical coordinates.}
The first color matrix on the right-hand side of \eqref{eq:shorttetra1} and \eqref{eq:shorttetra2} guarantees that the state projects onto a heavy quark-antiquark pair in a color octet configuration. 
Moreover, having made explicit the light quark content constrains the state to become an adjoint meson at short distance, and a tetraquark once combined with the heavy quark-antiquark pair.

The projection operator $\mathcal{P}_{\chi_{c1}(3872)}$ projects onto states made of an unpolarized $\chi_{c1}(3872)$ at rest ($\bm{P}=\bm{0}$) plus other light degrees of freedom generically denoted by $X$:
\begin{align}
\mathcal{P}_{\chi_{c1}(3872)} &= \sum_{X,m_J}\ket{\chi_{c1}(3872) + X} \bra{\chi_{c1}(3872) + X}.
\label{eq:PX3872}
\end{align}
Similarly to~\eqref{eq:X3872state}, we can approximate these states as 
\begin{align}
&     \ket{\chi_{c1}(3872) + X} \approx 
\sum_{m_l,\,m_S}\,C_{J=1,\,m_J;\,l=1,\,S=1}^{\,m_l,\,m_S}
     \int d^3R\,d^3r\, \bigg\{ \nonumber\\
     & \hspace{10mm} \psi_{\alpha,i}^{\dagger}(\bm{R} + \bm{r}/2)\chi_{\beta, j}(\bm{R} -\bm{r}/2)
     \bigg[(\Psi_{\Sigma_g^{+\prime}}^{\lambda=0,X})_{\alpha\beta}^{l=1,m_l;S=1,m_S}(\bm{r})\ket{X;1^{--}, \Sigma_g^{+\prime}(\lambda=0); \bm{r}; i,j} \nonumber\\
     &\hspace{23mm} +\frac{1}{\sqrt{2}}\bigg((\Psi_{\Pi_g}^{\lambda=1,X})_{\alpha\beta}^{l=1,m_l;S=1,m_S}(\bm{r})\ket{X;1^{--}, \Pi_g(\lambda=1); \bm{r}; i,j}\nonumber\\
     &\hspace{33mm} +(\Psi_{\Pi_g}^{\lambda=-1,X})_{\alpha\beta}^{l=1,m_l;S=1,m_S}(\bm{r})\ket{X;1^{--}, \Pi_g(\lambda=-1); \bm{r}; i,j}\bigg)\bigg] \bigg\},   
     \label{eq:X3872+Xstate}  
\end{align}
where we have neglected the quarkonium component of the $\chi_{c1}(3872)$, which is suppressed. 
The sum in \eqref{eq:PX3872} runs over all final state particles $X$ made out of light degrees of freedom and over the polarizations of the $\chi_{c1}(3872)$.
At the origin, the states $|X;1^{--}, \Sigma_g^{+\prime}(\lambda=0);$ $\bm{r}; i,j\rangle$ and $\ket{X;1^{--}, \Pi_g(\lambda=\pm 1); \bm{r}; i,j}$
behave similarly to eqs.~\eqref{eq:shorttetra1} and~\eqref{eq:shorttetra2}, but with the vacuum replaced by $\ket{X}$, 
which is a generic state made of light degrees of freedom:
\begin{align}
\ket{X;1^{--},\Sigma_g^{+\prime}(\lambda=0);\bm{0};i,j} &= \nonumber\\
& \hspace{-18mm} \left(\frac{2}{\left|\bar{q}(\bm{0})  \bm{P}_{10} \cdot \bm{\gamma} \, T^B q(\bm{0})\ket{X}\right|^2}\right)^{1/2} T^A_{ij} \bar{q}(\bm{R}) \bm{P}_{10} \cdot \bm{\gamma} \, T^A q(\bm{R}) \ket{X},
\label{eq:shorttetraX1}\\
\ket{X;1^{--},\Pi_g(\lambda=\pm 1);\bm{0};i,j} &= \nonumber\\
& \hspace{-18mm} \left(\frac{2}{\left| \bar{q}(\bm{0})  \bm{P}_{1\pm 1} \cdot \bm{\gamma} \, T^B q(\bm{0})\ket{X}\right|^2}\right)^{1/2} T^A_{ij} \bar{q}(\bm{R}) \bm{P}_{1\pm 1} \cdot \bm{\gamma} \, T^A q(\bm{R}) \ket{X}.
\label{eq:shorttetraX2}
\end{align}
The corresponding wave functions are
\begin{equation}
    (\Psi_{\Lambda}^{\lambda,X})^{l=1,m_l;S=1,m_S}_{\alpha\beta}(\bm{r}) = 
    v_{l=1,m_l}^{\lambda}\,\phi_{\Lambda}^{(X)}(r) \,\frac{{\bm e}_{m_S}\cdot \bm{\sigma}_{\alpha\beta}}{\sqrt{2}}.
    \label{eq:wavefctX3872}
\end{equation}

Inserting the projector into \eqref{eq:OperatorX3872}, the unpolarized octet LDME becomes\
\begin{align}
&\bra{\Omega}\mathcal{O}^{\chi_{c1}(3872)}({}^3S_1^{[8]})\ket{\Omega} = 
2\times \frac{3}{4\pi}\sum_X \bigg\{
\left| \phi^{(X)}_{\Sigma_g^{+\prime}}(0) \right|^2 \bra{\Omega} \Phi^{\dagger AB}_\ell T^A_{ji}  \ket{X;1^{--}, \Sigma_g^{+\prime}(\lambda=0); \bm{0} ;i,j}
\nonumber\\
&\hspace{70mm} \times\bra{X;1^{--}, \Sigma_g^{+\prime}(\lambda=0); \bm{0};j^\prime,i^\prime} T^C_{i'j'} \Phi^{BC}_\ell \ket{\Omega}
\nonumber\\
&\hspace{30mm}  + \left| \phi^{(X)}_{\Pi_g}(0) \right|^2 \frac{1}{2} \sum_{\lambda=\pm1}
\bra{\Omega} \Phi^{\dagger AB}_\ell T^A_{ji}  \ket{X;1^{--}, \Pi_g(\lambda); \bm{0} ;i,j}
\nonumber\\
&\hspace{60mm}\times \bra{X;1^{--}, \Pi_g(\lambda); \bm{0};j^\prime,i^\prime} T^C_{i'j'} \Phi^{BC}_\ell \ket{\Omega}\bigg\},
\label{octetpNRQCD}
\end{align}
where we have used anticommutation relations and $\psi\ket{\Omega} = \chi^\dagger\ket{\Omega} = 0$ to eliminate the heavy quark and antiquark fields.
As a result, the space integrals have been computed over the delta functions, and the wave functions are located at the origin.
The first factor 2 comes from $\Tr\{\sigma^k {\bm e}_{m_S}\cdot \bm{\sigma} \}/\sqrt{2} \times \Tr\{\sigma^k {\bm e}_{m_S'}\cdot \bm{\sigma} \}/\sqrt{2} = 2 \,\delta_{m_Sm_S'}$.
Then the sum over the Clebsch--Gordan coefficients leads to $\displaystyle \sum_{m_J,m_S} C_{J=1,\,m_J;\,l=1,\,S=1}^{\,m_l,\,m_S} C_{J=1,\,m_J;\,l=1,\,S=1}^{\,m_l',\,m_S}  = \delta_{m_lm_l'}$,
which is a special case of 
\begin{equation}
\sum_{m_J,m_S} C_{J,\,m_J;\,l,\,S}^{\,m_l,\,m_S}C_{J,\,m_J;\,l',\,S}^{\,m_l',\,m_S} = \delta_{ll'}\,\delta_{m_lm_l'}\,\frac{2J+1}{2l+1},
\label{eq:ClGo}
\end{equation}
and the sum over $m_l$ of the angular wave functions gives $\displaystyle \sum_{m_l} v_{l=1,m_l}^{\lambda'*}  v_{l=1,m_l}^{\lambda} = 3\delta_{\lambda\lambda'}/(4\pi)$, which is a special case of~\cite{Berwein:2015vca}
\begin{equation}
\sum_{m_l} v_{l,m_l}^{\lambda'*}  v_{l,m_l}^{\lambda} = \frac{2l+1}{4\pi}\delta_{\lambda\lambda'}\,.
\label{eq:vl}
\end{equation}
This last sum provides the factor $3/(4\pi)$ and an orthogonality relation between tetraquark states with different polarization. 
Equation~\eqref{octetpNRQCD} factorizes the heavy quark dependence entirely in the radial part of the wave functions at the origin, 
while the matrix elements depend  only on the light degrees of freedom.
The pNRQCD factorization of the NRQCD LDME is of advantage with respect to the NRQCD expression if it is possible to compute the heavy flavor dependent part of \eqref{octetpNRQCD}. 
If this is possible, then the low-energy part may be fitted to data and eventually used 
to estimate other production cross sections, most obviously the ones involving tetraquarks in the bottomonium sector.

The heavy quark dependent part of \eqref{octetpNRQCD} is encoded in the radial wave functions $\phi^{(X)}_{\Lambda}(r)$ evaluated at the origin.
They describe the different tetraquark components of states made of a $\chi_{c1}(3872)$ and some light degrees of freedom. 
These wave functions are so far unknown.
Henceforth, to proceed further, we follow a reasoning similar to the one used for quarkonium production in~\cite{Brambilla:2020ojz,Brambilla:2021abf,Brambilla:2022rjd,Brambilla:2022ayc}, 
i.e., we assume that the addition of final state light particles to the $\chi_{c1}(3872)$ does not modify its wave function significantly, as it leads to a constant shift in the tetraquark potential up to corrections that are suppressed in the large $N_c$ limit.
The ensuing approximation is 
\begin{equation}
\phi^{(X)}_{\Lambda}(r) \approx \phi_{\Lambda}(r)\,.
\label{approxlargeNc}
\end{equation}
The wave functions $\phi_{\Lambda}(r)$ are known from the solution of \eqref{coupledI0}, see~\cite{Berwein:2024ztx,Brambilla:2024imu}.
The octet LDME can then be written as 
\begin{equation}
\bra{\Omega}\mathcal{O}^{\chi_{c1}(3872)}({}^3S_1^{[8]})\ket{\Omega} = 
\frac{3}{2\pi} \left[ \left|\phi_{\Sigma_g^{+\prime}}(0) \right|^2 \mathcal{M}_{\Sigma_g^{+\prime}} 
+ \left|\phi_{\Pi_g}(0) \right|^2 \mathcal{M}_{\Pi_g} \right],
\label{octetpNRQCD2}
\end{equation}
where
\begin{align}
\mathcal{M}_{\Sigma_g^{+\prime}} &= \sum_X \left|\bra{\Omega} \Phi^{\dagger AB}_\ell T^A_{ji}  \ket{X;1^{--}, \Sigma_g^{+\prime}(\lambda=0); \bm{0} ;i,j}  \right|^2,
\label{MSigmagprime}\\
\mathcal{M}_{\Pi_g} &= \sum_{X} \, \frac{1}{2} \sum_{\lambda=\pm1} \left|\bra{\Omega} \Phi^{\dagger AB}_\ell T^A_{ji}  \ket{X;1^{--}, \Pi_g(\lambda); \bm{0} ;i,j}  \right|^2.
\label{MPig}
\end{align}

Expressed in matrix form, the octet LDME reads
\begin{equation}
\bra{\Omega}\mathcal{O}^{\chi_{c1}(3872)}({}^3S_1^{[8]})\ket{\Omega} = \frac{3}{2\pi} \,  
\left(\phi^\dagger_{\Sigma_g^{+\prime}}(0), \phi^\dagger_{\Pi_g}(0)\right)
\begin{pmatrix}
 \mathcal{M}_{\Sigma_g^{+\prime}} &   0\\
  0 & \mathcal{M}_{\Pi_g}
\end{pmatrix}
\begin{pmatrix}
\phi_{\Sigma_g^{+\prime}}(0) \\
\phi_{\Pi_g}(0) 
\end{pmatrix},
\label{octetpNRQCD3}
\end{equation}
which makes it evident that the matrix element is invariant under a simultaneous unitary rotation of the wave functions and the matrix $\mathcal M$.
A convenient unitary transformation is 
\begin{equation}
    U = \frac{1}{\sqrt{3}}
\begin{pmatrix}
  -1& -\sqrt{2} \\
  -\sqrt{2} & 1 
\end{pmatrix} 
= U^{-1}.
\end{equation}
This transformation makes the {\it centrifugal barrier} in eq.~\eqref{coupledI0} diagonal:\footnote{
We focus on the tetraquark sector here. 
The transformation may be extended to the quarkonium sector by taking
\begin{equation*}
    U = \frac{1}{\sqrt{3}}
\begin{pmatrix}
   \sqrt{3} &0&0\\
  0&-1& -\sqrt{2} \\
  0&-\sqrt{2} & 1 
\end{pmatrix}.
\end{equation*}
}
\begin{equation}
U  {\begin{pmatrix}
 4        & - 2\sqrt{2} \\
 -2 \sqrt{2} & 2
\end{pmatrix}} U^{-1} = 
{\begin{pmatrix}
 0        & 0 \\
 0 & 6
\end{pmatrix}}.
\label{Ukinetic}
\end{equation}
From the diagonalized centrifugal barrier, it follows that we can identify the transformed wave functions with an $S$-wave function, $\phi_S$, and a $D$-wave function, $\phi_D$:\footnote{
For a system made of two particles of mass $m$, the centrifugal barrier has the form $l(l+1)/(mr^2)$
on an eigenstate of the orbital angular momentum with quantum number $l$. 
}
\begin{equation}
U \begin{pmatrix}
 \phi_{\Sigma_g^{+\prime}} \\
\phi_{\Pi_g}  
\end{pmatrix} \equiv
\begin{pmatrix}
 \phi_{S}\\
 \phi_{D}
\end{pmatrix}.
\label{eq:Uphi}
\end{equation}
Finally, under the transformation $U$ the matrix $\mathcal{M}$ transforms into 
\begin{equation}
  U \begin{pmatrix}
 \mathcal{M}_{\Sigma_g^{+\prime}} &  0 \\
 0 &\mathcal{M}_{\Pi_g} 
\end{pmatrix} U^{-1} \equiv
\begin{pmatrix}
 \mathcal{M}_{S} & \mathcal{M}_{SD} \\
 \mathcal{M}_{SD}^\dagger &\mathcal{M}_{D} 
\end{pmatrix},
\end{equation}
with
\begin{align}
\mathcal{M}_{S} &= \frac{1}{3} \left( \mathcal{M}_{\Sigma_g^{+\prime}} + 2 \mathcal{M}_{\Pi_g} \right),
\label{MS}\\
\mathcal{M}_{D} &= \frac{1}{3} \left(  2 \mathcal{M}_{\Sigma_g^{+\prime}}+\mathcal{M}_{\Pi_g} \right),
\label{MD}\\
\mathcal{M}_{SD} &=  \frac{\sqrt{2}}{3} \left(\mathcal{M}_{\Sigma_g^{+\prime}}- \mathcal{M}_{\Pi_g}  \right).
\label{MSD}
\end{align}
In a different language, the transformation $U$ changes from an {\it adiabatic basis}, where the potential matrix in \eqref{coupledI0} is diagonal in the tetraquark sector, to a {\it diabatic basis}, where the kinetic energy matrix in \eqref{coupledI0} is diagonal.
In the diabatic basis, using a charm mass of about $1.4$~GeV,\footnote{
\label{footnotemass}
The charm mass of about $1.4$~GeV should be properly identified with the so-called Cornell mass~\cite{Mateu:2018zym} 
(in the context of the computation of NRQCD LDMEs see also~\cite{Bodwin:2007fz}).
If we use the parameterization of the potentials given in~\cite{Brambilla:2024imu}, which reflects the lattice parameterization of~\cite{Bulava:2024jpj}, then a Cornell mass of about $1.4$~GeV corresponds to a mass of about $1.97$~GeV to be used in the coupled Schr\"odinger equations \eqref{coupledI0}. 
Twice a charm mass of about $1.97$~GeV matches, indeed, the value of the long-distance
spin-averaged meson-antimeson threshold set in the lattice parameterization.
In this section and the following one, we use $1.4$~GeV also for the charm mass to be used in the short-distance coefficients.}
we obtain for the wave functions at the origin, solutions of the coupled Schr\"odinger equations~\eqref{coupledI0},
\begin{align}
 |\phi_{\Sigma_g^+}(0)|^2  &\approx 0,\label{cwavefunction0}\\
 |\phi_{S}(0)|^2 &= 5.78 \times 10^{-3}  \, \mathrm{GeV}^3,\label{cwavefunction}\\
 |\phi_{D}(0)|^2 &\approx 0, \label{cwavefunction0D}
\end{align}
where we have added the quarkonium component, which is unaffected by the unitary rotation.
We see that the change of basis is convenient because it singles out one component, the $S$-wave tetraquark component, that is largely dominant over the other $\chi_{c1}(3872)$ components.\footnote{
The suppression of the quarkonium component can be understood as the vanishing at the origin of a $P$-wave function.
}
This leads to the final pNRQCD factorization for the unpolarized octet LDME:
\begin{equation}
\bra{\Omega}\mathcal{O}^{\chi_{c1}(3872)}({}^3S_1^{[8]})\ket{\Omega} = 
\frac{3}{2\pi} \, \left|\phi_{S}(0)\right|^2 \,\mathcal{M}_{S},
\label{octetpNRQCD4}
\end{equation}
with $\mathcal{M}_{S}$ defined by eqs. \eqref{MS}, \eqref{MSigmagprime} and \eqref{MPig}:
\begin{align} 
\mathcal{M}_{S} = \frac{1}{3} \sum_X & \bigg(  \left|\bra{\Omega} \Phi^{\dagger AB}_\ell T^A_{ji}  \ket{X;1^{--}, \Sigma_g^{+\prime}(\lambda=0); \bm{0} ;i,j}  \right|^2 
\nonumber\\
& +\sum_{\lambda=\pm1} \left|\bra{\Omega} \Phi^{\dagger AB}_\ell T^A_{ji}  \ket{X;1^{--}, \Pi_g(\lambda); \bm{0} ;i,j}  \right|^2\bigg).
\label{MS2}
\end{align}
The absence of a gauge field correlator, unlike in the quarkonium case \eqref{eq:O8pNRQCD}, reflects at the level of pNRQCD the fact that the octet matrix element contributes to the production of the $\chi_{c1}(3872)$ at leading order in $v^2$.

In the bottomonium sector, the equivalent of the $\chi_{c1}(3872)$ state is called $X_b$~\cite{Brambilla:2024imu}.
The production cross section of this state is described in the same way as the production cross section of the $\chi_{c1}(3872)$, and the octet LDME can be factorized as in \eqref{octetpNRQCD4}.
The factor $\mathcal{M}_S$ is universal and the same as for the $\chi_{c1}(3872)$ production.
The difference lies in the wave function, which depends on the mass of the heavy quark.
For a bottom mass of about 4.74~GeV,\footnote{
As in the charm case, we use this mass in this section and in the following one for the short-distance coefficients as well.}
the wave function at the origin turns out to be
\begin{equation}
|\phi_{S}(0)|^2 \approx 0.158 \,\mathrm{GeV}^3.
\label{bwavefunction}
\end{equation}

A similar derivation of the inclusive $\chi_{c1}(3872)$ cross section in the framework of pNRQCD, as the one presented here, was first done in~\cite{Lai:2025tpw}. 
Our final factorization of the octet LDME, eq.~\eqref{octetpNRQCD4}, agrees formally with the factorization obtained there.
However, the explicit expression of the universal factor $\mathcal{M}_S$, which we give in eq. \eqref{MS2} is different from the one there.
The origin of the difference can be traced back to the sum of intermediate states in the projector operator \eqref{eq:PX3872}. 
While in our case we restrict the sum to states containing a $\chi_{c1}(3872)$, i.e. a heavy quark-antiquark pair in a color octet configuration and a light quark pair forming an adjoint meson with quantum numbers $1^{--}$ at short distance, in~\cite{Lai:2025tpw} the sum is taken over all states containing a color octet heavy quark-antiquark pair: these states include quarkonium hybrids, pentaquarks and possibly other exotic quarkonium states. 
By approximating the wave functions as in \eqref{approxlargeNc}, 
the extended sum in~\cite{Lai:2025tpw} eventually leads to a completion condition on the sum of states that allows to approximate the factor $\mathcal{M}_S$ by $(1/2)(N_c^2-1)/3 = 4/3$, the factor $1/2$ being the normalization of the SU($N_c$) generators.\footnote{
Further differences with~\cite{Lai:2025tpw} concerning $\mathcal{M}_S$ are in the factor $N_c^2-1$, which is absent in~\cite{Lai:2025tpw} possibly due to a different normalization, and in the factor $1/3$, which in our case follows from $|\phi_{\Sigma_g'}(0)|^2 \approx |\phi_S(0)|^2/3$
and $|\phi_{\Pi_g}(0)|^2 \approx 2 |\phi_S(0)|^2/3$, see eq.~\eqref{eq:Uphi}, while it is absent in~\cite{Lai:2025tpw} 
possibly due to having expressed the octet LDME in terms of $|\phi_{\Sigma_g'}(0)|^2$ rather than $|\phi_S(0)|^2$,
but no explicit distinction is made there between the different wave functions.}  
Since, instead, the sum is restricted to states containing  a $\chi_{c1}(3872)$ only, 
we rather interpret here $4/3$ as an upper limit on $\mathcal{M}_S$.
The upper limit on $\mathcal{M}_S$ leads to a non-trivial upper limit on the octet LDME:
\begin{equation}
\bra{\Omega}\mathcal{O}^{\chi_{c1}(3872)}({}^3S_1^{[8]})\ket{\Omega} \lesssim  
\frac{3}{2\pi} \times (5.78\times 10^{-3} \,\mathrm{GeV}^3) \times \frac{4}{3} \approx 3.68\times 10^{-3} \,\mathrm{GeV}^3.
\label{eq:OX3872upper}
\end{equation}
The upper limit relies on the positivity of each term in the sum of $\mathcal{M}_S$. 
This is guaranteed as long as the LDME does not undergo renormalization, which is the case at the accuracy of the present calculation.

\subsection{Production of $\chi_{c1}(3872)$ from $B$-decay}
\label{sec:X3872Bdecay}
In the framework of the NRQCD factorization, the octet LDME $\bra{\Omega}\mathcal{O}^{\chi_{c1}(3872)}({}^3S_1^{[8]})\ket{\Omega}$ 
can be extracted from the $\chi_{c1}(3872)$ formation branching ratio (Br) in $B$ decays.
At leading order in the velocity expansion and $\Lambda_{\rm QCD}/m_{c,b}$, 
the branching ratio depends on the same octet LDME contributing to the $\chi_{c1}(3872)$ hadroproduction cross section, namely $\bra{\Omega}\mathcal{O}^{\chi_{c1}(3872)}({}^3S_1^{[8]})\ket{\Omega}$.
It is given by~\cite{Beneke:1998ks}
\begin{align}
    \mathrm{Br}(B\rightarrow \chi_{c1}(3872) +X) &= \mathrm{Br}(b \rightarrow c\bar{c}({}^3S_1^{[8]})+X)\,\bra{\Omega}\mathcal{O}^{\chi_{c1}(3872)}({}^3S_1^{[8]})\ket{\Omega}.
 \label{eq:BX3872}   
\end{align}
The short-distance coefficient $\mathrm{Br}(b \rightarrow c\bar{c}({}^3S_1^{[8]})+X)$ is known at next-to-leading order in the strong coupling from~\cite{Beneke:1998ks}. 
We set the bottom mass to $m_b=4.74 \, \mathrm{GeV}$, the charm mass to $m_c=1.4 \, \mathrm{GeV}$,
the renormalization scale to $\mu = m_b$ and use the two-loop running formula with $\Lambda_{\overline{\mathrm{MS}}}^{(5)} = 226 \, \mathrm{MeV}$ for $n_f=5$ active flavors for $\alpha_s(m_b)$. 
We thus obtain $\mbox{Br}(b \rightarrow \bar{c}c({}^3S_1^{[8]})+X) = 0.213^{+0.019}_{-0.028} \, \mathrm{GeV}^{-3}$, where the error is estimated by varying the renormalization scale in the range from $\mu  = m_b/2$ to $\mu = 2m_b$.
We determine the left-hand side of \eqref{eq:BX3872} from the LHCb measurement of $\mathrm{Br}(B\to \chi_{c1}(3872) + X)$ $\times$  $\mathrm{Br}(\chi_{c1}(3872)$ $\to$  $J/\psi \pi^+\pi^-)=(4.3 \, \pm \, 0.5)\times 10^{-5}$ \cite{LHCb:2021ten} 
and the PDG value for $\mathrm{Br}(\chi_{c1}(3872)$ $\to J/\psi \pi^+\pi^-) = (4.3 \pm 1.4)\%$ \cite{ParticleDataGroup:2024cfk},  
which give $\mathrm{Br}(B \rightarrow \chi_{c1}(3872) + X) = (1.0 \pm 0.3  )\times 10^{-3}$. 
On top of the given uncertainties in the experimental value and the theoretical short-distance coefficient of the $B$ decay, there is an error of $\mathcal{O}(v^2)$ due to higher order corrections in \eqref{eq:BX3872}, which we estimate to be 30\%. 
Hence, the errors are given by the scale uncertainties of the short-distance coefficient, the experimental uncertainties of the branching ratio, and the error due to higher-order terms in $v$, which are added in quadrature.
This leads to the following numerical value for the octet matrix element
\begin{equation}
    \bra{\Omega} \mathcal{O}^{\chi_{c1}(3872)}({}^3S_1^{[8]})\ket{\Omega} = 4.69^{+2.23}_{-2.16} \times 10^{-3} \, \mathrm{GeV}^3,
\label{eq:NumericalOX3872}
\end{equation}    
and consequently, using \eqref{cwavefunction}, for the pNRQCD matrix element 
\begin{equation}
    \mathcal{M}_S = 1.70^{+0.81}_{-0.78} \,.
    \label{eq:NumericalM}
\end{equation}

\begin{figure}[ht]
    \centering 
    \begin{subfigure}{0.48\textwidth}
        \includegraphics[scale=0.36]{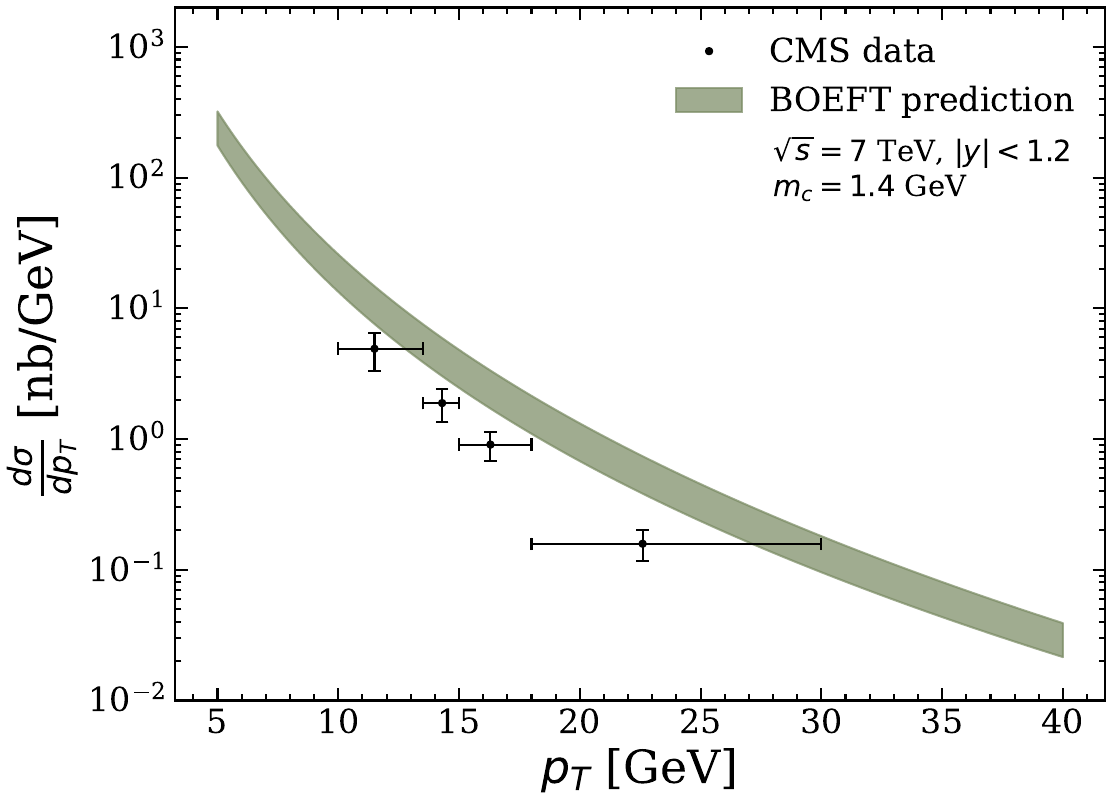}
    \end{subfigure}
    \begin{subfigure}{0.48\textwidth}
        \includegraphics[scale=0.36]{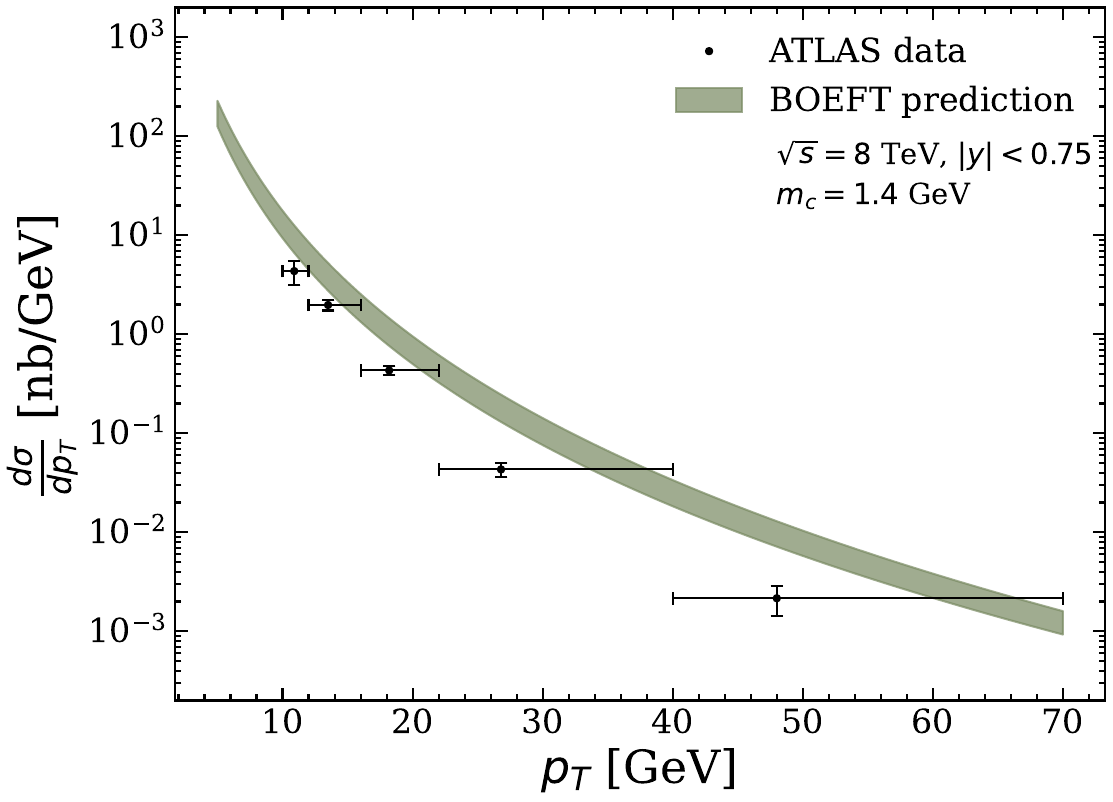}
    \end{subfigure}
    \\
    \begin{subfigure}{0.48\textwidth}
        \includegraphics[scale=0.36]{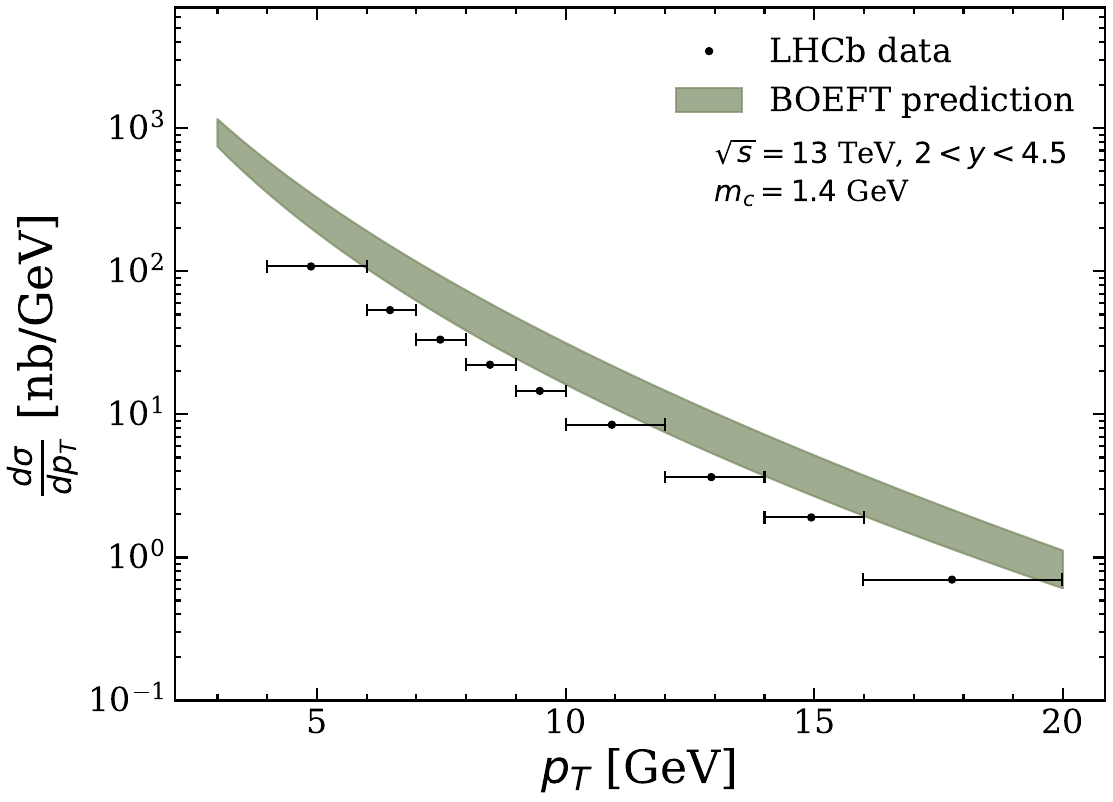}
    \end{subfigure}
    \caption{Theoretical predictions for the prompt inclusive hadroproduction differential cross section of the $\chi_{c1}(3872)$ state compared with CMS data~\cite{CMS:2013fpt} (upper-left panel),   
    ATLAS data~\cite{ATLAS:2016kwu} (upper-right panel) and LHCb data~\cite{LHCb:2021ten} (lower panel).
    The central values of the CMS and ATLAS data are taken from~\cite{hepdata.60421} and~\cite{hepdata.76839}, respectively;
    the central values of the LHCb data have been computed following~\cite{Lafferty:1994cj}.
    The region above the upper bound \eqref{eq:OX3872upper} has been excluded from the bands in the plots, 
    effectively reducing the uncertainties of the NRQCD predictions.}
    \label{fig:SigmaXc}
\end{figure}

\subsection{Hadroproduction of $\chi_{c1}(3872)$\label{sec:HadroXc}}
Inserting the value of the octet LDME \eqref{eq:NumericalOX3872} in eq.~\eqref{eq:NRQCDfacX3872}, 
we can predict the prompt unpolarized inclusive hadroproduction cross section of the $\chi_{c1}(3872)$. 
Here and in the following sections, we calculate the short distance coefficients at next-to-leading order in $\alpha_s$, using the codes of ref.~\cite{Butenschoen:2010rq}. 
We thereby choose the charm and bottom masses to be $m_c=1.4$~GeV and $m_b=4.74$~GeV. 
We set the renormalization and factorization scales to $\mu_r=\mu_f=\sqrt{p_T^2+m_Q^2}$, use the CTEQ6M~\cite{Pumplin:2002vw} proton parton distribution function set, and, correspondingly, use the two loop running formula for $\alpha_s(\mu_r)$ with $\Lambda^{(4)}_{\overline{\mathrm{MS}}} = 326$~MeV for charmed and $\Lambda^{(5)}_{\overline{\mathrm{MS}}}= 226$~MeV for bottomed hadrons. 
The predicted differential cross section is shown in figure~\ref{fig:SigmaXc} compared with CMS, ATLAS, and LHCb data. 
Error bands are obtained by adding in quadrature the uncertainty of $\mathcal{M}_S$ and the scale uncertainty of the short-distance coefficient, where the latter is estimated by varying the factorization and renormalization scales simultaneously by a factor $2$ and $1/2$.
Moreover, the BOEFT factorization formula \eqref{octetpNRQCD4} supplements the NRQCD result with the upper bound \eqref{eq:OX3872upper}.
This upper bound excludes certain regions from the NRQCD prediction, thereby significantly reducing its uncertainties. 
The error bands shown in figure~\ref{fig:SigmaXc} combine the NRQCD prediction with the BOEFT upper bound. 
It should be noted that the non trivial fact that the upper bound $\mathcal{M}_S \le 4/3$ falls inside the range \eqref{eq:NumericalM} may indicate that the sum over intermediate octet states in the production LDME is dominated by the production of the $\chi_{c1}(3872)$.

\subsection{Hadroproduction of $X_b$}
\begin{figure}[ht]
    \centering
    \includegraphics[scale=0.4]{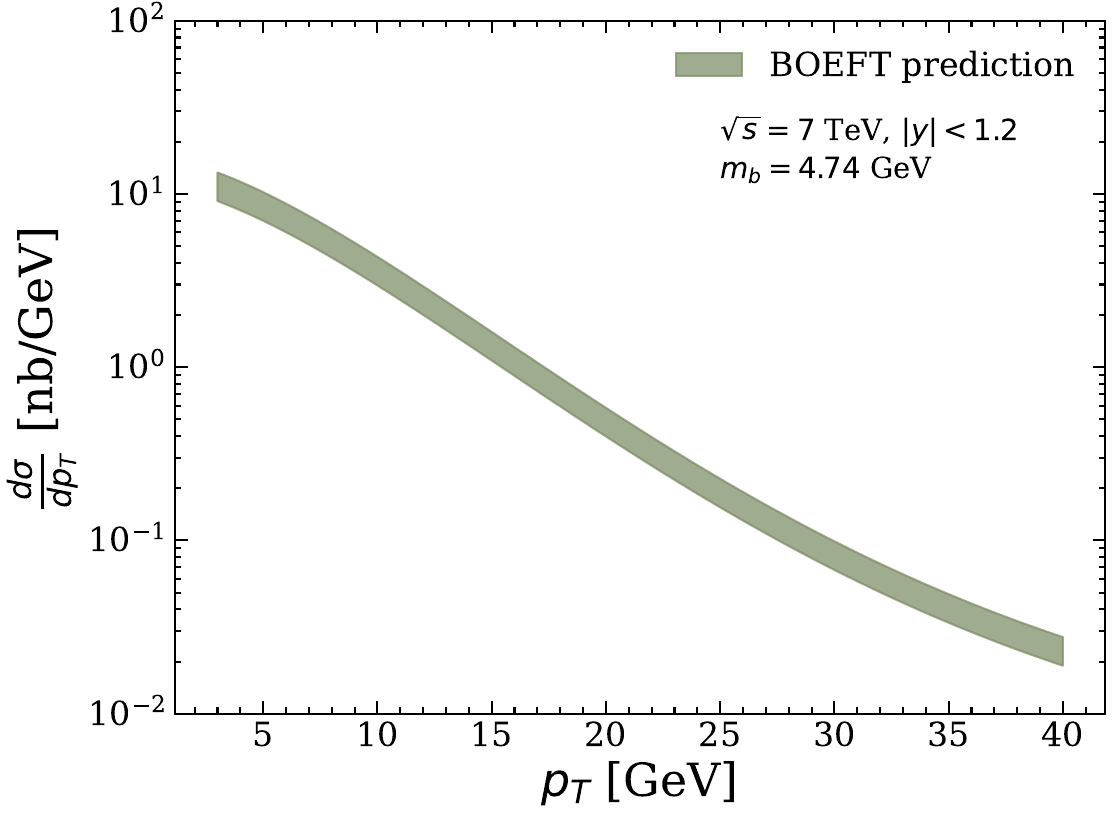}
    \caption{Prediction for the prompt inclusive hadroproduction differential cross section of the $X_b$ using the CMS kinematics from the $\chi_{c1}(3872)$ production 
    and the value $m_b = 4.74\, \mathrm{GeV}$ for the bottom mass. 
    The region above the upper bound following from \eqref{eq:OX3872upper} has been excluded from the band in the plot.}
    \label{fig:Xb_cross_section}
\end{figure}

Once the universal matrix element $\mathcal{M}_S$ has been established from the $\chi_{c1}(3872)$ octet LDME, 
see eq.~\eqref{eq:NumericalM}, it can be used to make predictions in the bottomonium sector as well.
In the bottomonium sector, the tetraquark state with the same quantum numbers as the $\chi_{c1}(3872)$ is named $X_b$.
Solving the coupled Schr\"odinger equations \eqref{coupledI0} for the bottom mass, we obtain the wave function at the origin \eqref{bwavefunction}.
Combining it with $\mathcal{M}_S$ given in \eqref{eq:NumericalM}, we predict the prompt inclusive hadroproduction differential cross section of the~$X_b$.
The result with the same CMS kinematics as for the $\chi_{c1}(3872)$ production is displayed in figure~\ref{fig:Xb_cross_section}; 
errors are estimated as in the $\chi_{c1}(3872)$ case.
This is a genuine pNRQCD prediction that follows from the factorization formula~\eqref{octetpNRQCD4}.

\section{Inclusive production of pentaquarks}
\label{sec:QQbarqqq}
Pentaquark states in the charmonium spectrum were discovered ten years ago at the LHCb~\cite{LHCb:2015yax}.
In this section, we address the hadroproduction of the pentaquark states $P_{c\bar{c}}(4312)^+$, $P_{c\bar{c}}(4457)^+$, $P_{c\bar{c}}(4380)^+$, and $P_{c\bar{c}}(4440)^+$ in the BOEFT framework 
developed above for the $\chi_{c1}(3872)$.
We do so in the two distinct scenarios analyzed in~\cite{Brambilla:2025xma} ({\it scenario I}) and~\cite{Alasiri:2025roh} ({\it scenario II}).
In~\cite{Brambilla:2025xma}, it is assumed that all three Born--Oppenheimer potentials for pentaquarks labeled $(1/2)_g$,  $(1/2)'_g$ and  $(3/2)_g$ support bound states.\footnote{
Born--Oppenheimer potentials for pentaquarks are classified in terms of the total angular momentum of the light quarks projected on the heavy quark-antiquark axis and parity ($g$ stands for positive parity).
The Born--Oppenheimer potential $(1/2)_g$ joins at short distance with the mass of a $(1/2)^+$ adjoint baryon. 
The Born--Oppenheimer potentials $(1/2)'_g$ and  $(3/2)_g$ are degenerate at short distance with the mass of a $(3/2)^+$ 
adjoint baryon.}
This leads to assigning the $J^P$ quantum numbers $(1/2)^-$ to $P_{c\bar{c}}(4312)^+$ and $P_{c\bar{c}}(4457)^+$, and the $J^P$ quantum numbers $(3/2)^-$ to $P_{c\bar{c}}(4380)^+$ and $P_{c\bar{c}}(4440)^+$.
In contrast, in~\cite{Alasiri:2025roh}, it is assumed that only the two Born--Oppenheimer potentials labeled $(1/2)'_g$ and  $(3/2)_g$ support bound states.
This leads to assign the $J^P$ quantum numbers $(1/2)^-$ to $P_{c\bar{c}}(4312)^+$, the $J^P$ quantum numbers $(5/2)^-$ to $P_{c\bar{c}}(4457)^+$, and 
the $J^P$ quantum numbers $(3/2)^-$ to $P_{c\bar{c}}(4380)^+$ and $P_{c\bar{c}}(4440)^+$.
At present, both scenarios are possible as neither the Born--Oppenheimer potentials for pentaquarks in lattice QCD nor the pentaquark quantum numbers are known.

\subsection{Hadroproduction of charmonium pentaquarks in scenario I}
In scenario I, the pentaquark states $P_{c\bar{c}}(4312)^+$, $P_{c\bar{c}}(4457)^+$, $P_{c\bar{c}}(4380)^+$ and $P_{c\bar{c}}(4440)^+$ are written as~\cite{Brambilla:2025xma} 
\begin{align}
    \ket{P_{c\bar{c}}(4312)^+} &= a_1\ket{S=0;k = 1/2}_{J^P=(1/2)^-} + a_2\ket{S=1;k = 1/2}_{J^P=(1/2)^-} \nonumber\\
    & \quad + a_3\ket{S=1;k = 3/2}_{J^P=(1/2)^-}
    \label{eq:PentaState4312} \\
    \ket{P_{c\bar{c}}(4457)^+} &= b_1\ket{S=0;k = 1/2}_{J^P=(1/2)^-} + b_2\ket{S=1;k = 1/2}_{J^P=(1/2)^-} \nonumber\\
    & \quad + b_3\ket{S=1;k = 3/2}_{J^P=(1/2)^-}\,, \label{eq:PentaState4457}\\
    \ket{P_{c\bar{c}}(4380)^+} &= c_1\ket{S=1;k = 1/2}_{J^P=(3/2)^-} + c_2\ket{S=0;k = 3/2}_{J^P=(3/2)^-} \nonumber\\
    & \quad + c_3\ket{S=1;k = 3/2}_{J^P=(3/2)^-}\,, \label{eq:PentaState4380} \\
    \ket{P_{c\bar{c}}(4440)^+} &= d_1\ket{S=1;k = 1/2}_{J^P=(3/2)^-} + d_2\ket{S=0;k = 3/2}_{J^P=(3/2)^-} \nonumber\\
    & \quad + d_3\ket{S=1;k = 3/2}_{J^P=(3/2)^-}\,, \label{eq:PentaState4440} 
\end{align}
with $a_1=0.475$, $a_2 =  0.267$, $a_3=0.838$, $b_1 = -0.331$, $b_2=-0.829$, $b_3=0.452$,  $c_1 = 0.536$, $c_2=-0.506$, $c_3 = 0.676$, and $d_1 = -0.314$, $d_2 = 0.624$, $d_3 = 0.716$. 
The states $\ket{S;k}_{J^P}$ are eigenstates of the heavy-quark-antiquark spin and the total angular momentum of the light quarks, whose quantum numbers are $S$ and $k$, respectively.
The states have also definite $J^P$ quantum numbers.
In a generic reference frame, with center of mass momentum $\bm P$, they can be written as  
\begin{align}
&\ket{S;k = 1/2}_{J^P=(1/2)^-} = \nonumber\\
& \hspace{5mm} \sum_{m_l,m_S}C^{m_l,m_S}_{J=1/2,m_J;l=1/2,S} \int d^3R\, d^3r\,e^{i\bm P\cdot \bm R} \,\psi^\dagger_{\alpha,i}(\bm{R} + \bm{r}/2)\chi_{\beta,j}(\bm{R} - \bm{r}/2)\nonumber\\
&\hspace{5mm}\times  \frac{1}{\sqrt{2}} \bigg[\left( \Psi^{\lambda=1/2}_{(1/2)_g} \right)^{l=1/2,m_l;S,m_S}_{\alpha\beta }(\bm{r}) \ket{0;(1/2)^+, (1/2)_g(\lambda=1/2); \bm{r}; i,j}+ (\lambda \to -1/2)\bigg], \label{eq:BOEFTPenta12}
\end{align}
\begin{align}
&\ket{S=1;k = 1/2}_{J^P=(3/2)^-} = \nonumber\\
& \hspace{5mm} \sum_{m_l,m_S}C^{m_l,m_S}_{J=3/2,m_J;l=1/2,S=1} \int d^3R\, d^3r\,e^{i\bm P\cdot \bm R}\, \psi^\dagger_{\alpha,i}(\bm{R} + \bm{r}/2)\chi_{\beta,j}(\bm{R} - \bm{r}/2)\nonumber\\
&\hspace{5mm}\times \frac{1}{\sqrt{2}} \bigg[ \left( \Psi^{\lambda=1/2}_{(1/2)_g} \right)^{l=1/2,m_l;S=1,m_S}_{\alpha\beta }(\bm{r}) \ket{0;(1/2)^+, (1/2)_g(\lambda=1/2); \bm{r}; i,j}+ (\lambda \to - 1/2) \bigg], \label{eq:BOEFTPenta12a}
\end{align}
\begin{align}
&\ket{S=1;k = 3/2}_{J^P=(1/2)^-} = \nonumber\\
& \hspace{5mm} \sum_{m_l,m_S}C^{m_l,m_S}_{J=1/2,m_J;l=3/2,S=1} \int d^3R\, d^3r\,e^{i\bm P\cdot \bm R} \,\psi^\dagger_{\alpha,i}(\bm{R} + \bm{r}/2)\chi_{\beta,j}(\bm{R} - \bm{r}/2)\nonumber\\
&\hspace{5mm}\times \frac{1}{\sqrt{2}} \bigg[ \left( \Psi^{\lambda=1/2}_{(1/2)'_g} \right)^{l=3/2,m_l;S=1,m_S}_{\alpha\beta}(\bm{r}) \ket{0;(3/2)^+, (1/2)'_g(\lambda=1/2); \bm{r}; i,j} + (\lambda \to - 1/2) \nonumber\\
&\hspace{14mm} +\left( \Psi^{\lambda=3/2}_{(3/2)_g} \right)^{l=3/2,m_l;S=1,m_S}_{\alpha\beta}(\bm{r}) \ket{0;(3/2)^+, (3/2)_g(\lambda=3/2); \bm{r}; i,j} +(\lambda \to - 3/2)\bigg], \label{eq:BOEFTPenta32a}
\end{align}
\begin{align}
&\ket{S;k = 3/2}_{J^P=(3/2)^-} = \nonumber\\
& \hspace{5mm} \sum_{m_l,m_S}C^{m_l,m_S}_{J=3/2,m_J;l=3/2,S} \int d^3R\, d^3r\,e^{i\bm P\cdot \bm R}\,   \psi^\dagger_{i,\alpha}(\bm{R} + \bm{r}/2)\chi_{j,\beta}(\bm{R} - \bm{r}/2)\nonumber\\
&\hspace{5mm} \times \frac{1}{\sqrt{2}}  \bigg[ \left( \Psi^{\lambda=1/2}_{(1/2)'_g} \right)^{l=3/2,m_l;S,m_S}_{\alpha\beta} (\bm{r}) \ket{0;(3/2)^+, (1/2)'_g(\lambda=1/2); \bm{r}; i,j} + (\lambda \to -1/2) \nonumber\\
&\hspace{14mm} 
+ \left( \Psi^{\lambda=3/2}_{(3/2)_g} \right)^{l=3/2,m_l;S,m_S}_{\alpha\beta } (\bm{r}) \ket{0;(3/2)^+, (3/2)_g(\lambda=3/2); \bm{r}; i,j} + (\lambda \to -3/2)\bigg].
\label{eq:BOEFTPenta32}
\end{align}
The meaning and the labeling of the states $\ket{0;(1/2)^+, (1/2)_g(\lambda); \bm{r}; i,j}$,  $|0;(3/2)^+, (1/2)'_g$ $(\lambda);$ $\bm{r}; i,j\rangle$ and 
$|0;(3/2)^+,$ $(3/2)_g$ $(\lambda);$ $ \bm{r}; i,j\rangle$ is the same as the one of the tetraquark states introduced in the previous section (see eq. \eqref{eq:X3872state}) and so for the wave functions $\Psi_{\Lambda}^\lambda(\bm{r})$.
These can be written as 
\begin{equation}
    \left(\Psi_{\Lambda}^\lambda\right)^{l,m_l;S,m_S}_{ \alpha\beta}(\bm{r}) = 
    v_{l,m_l}^{\lambda}\, \phi_{\Lambda}(r) \,
    \left(\chi_{S,m_S}\right)_{\alpha\beta} \,,
\end{equation}
where the spin part $\left(\chi_{S,m_S}\right)_{\beta \alpha}$ reads 
\begin{align}
    (\chi_{S=0,m_S})_{\alpha\beta} = \frac{\delta_{m_S,0}}{\sqrt{2}}\delta_{\alpha\beta}
    \quad \text{or}\quad 
     (\chi_{S=1,m_S})_{\alpha\beta} = \frac{{\bm e}_{m_S}\cdot \bm{\sigma}_{\alpha\beta}}{\sqrt{2}},
\end{align}
$\bm{e}_{m_S}$ being the three unit vectors that identify the polarizations of the spin $S=1$ state.
In scenario I, all three pentaquark states are linear combinations of $S=0$ and $S=1$ heavy quark spin states.

Since the components of the $P_{c\bar c}(4312)^+$ and $P_{c\bar c}(4457)^+$ states are the same, 
apart from the numerical constants $a_i$ and $b_i$, the calculation of the LDMEs is identical for these two pentaquarks. 
We begin by considering the hadroproduction cross section for the $P_{c\bar c}(4312)^+$ state.
As for the $\chi_{c1}(3872)$, also pentaquark states are produced with the heavy quark-antiquark pair in a color octet configuration~\cite{Berwein:2024ztx}.
At leading order in the velocity expansion, the hadroproduction cross section of the $P_{c\bar c}(4312)^+$ state depends on two octet LDMEs.
These are 
\begin{align}
    \bra{\Omega}\mathcal{O}^{P_{c\bar{c}}(4312)}({}^1S_0^{[8]})\ket{\Omega} &=\bra{\Omega}\chi^\dagger  T^A\psi\Phi_\ell^{\dagger AB}{\cal P}_{{P_{c\bar{c}}}(4312)
    }\Phi_\ell^{BC}\psi^\dagger T^C\chi\ket{\Omega},
    \label{OPcc0}\\
    \bra{\Omega}\mathcal{O}^{P_{c\bar{c}}(4312)}({}^3S_1^{[8]})\ket{\Omega}&=\bra{\Omega}\chi^\dagger \sigma^k T^A\psi\Phi_\ell^{\dagger AB}{\cal P}_{P_{c\bar{c}}(4312)}\Phi_\ell^{BC}\psi^\dagger\sigma^kT^C\chi\ket{\Omega}.
    \label{OPcc1}
\end{align}
The projector ${\cal P}_{P_{c\bar{c}}(4312)}$ projects onto states containing a $P_{c\bar{c}}(4312)^+$ at rest and final state particles $X$ made out of light degrees of freedom:
\begin{align}
    \mathcal{P}_{P_{c\bar{c}}(4312)} &= \sum_{X, m_J}\ket{P_{c\bar{c}}(4312)^++X
    }\bra{P_{c\bar{c}}(4312)^++X
    }.
\end{align}
Because of the sum over the polarizations $m_J$, ${\cal P}_{P_{c\bar{c}}(4312)}$ projects onto an unpolarized $P_{c\bar{c}}(4312)^+$.
The fact that the hadroproduction cross section of the $P_{c\bar{c}}(4312)^+$ involves two LDMEs follows 
from the pentaquark state being a linear combination of spin singlet and spin triplet heavy quark-antiquark pairs.
The same holds true for the other pentaquark states as well, see eqs.~\eqref{eq:PentaState4312}-\eqref{eq:PentaState4440}.

As in the $\chi_{c1}(3872)$ case, we assume that the wave functions $\left(\Psi_{\Lambda}^\lambda\right)^{l,m_l;S,m_S}_{\alpha\beta}(\bm{r})$ are approximately unaffected by the extra production of light states $X$, 
while the state $|0;k^+,\Lambda(\lambda);$ $\bm{r};i,j\rangle$ changes into $\ket{X;k^+,\Lambda(\lambda);\bm{r};i,j}$. 
The states $\ket{X;k^+,\Lambda(\lambda);\bm{r};i,j}$ are in general unknown. 
What is known is that at $\bm{r}=\bm{0}$ they contain three light quarks in a color octet configuration.
Once combined with the heavy quark-antiquark pair in a color octet configuration, 
this guarantees that a heavy pentaquark is produced.
To compute the octet LDMEs, we insert the explicit expression of the projector into eqs. \eqref{OPcc0} and \eqref{OPcc1}, 
express the state $\ket{P_{c\bar{c}}(4312)^++X}$ in terms of the states \eqref{eq:BOEFTPenta12} and \eqref{eq:BOEFTPenta32a} according to the above approximation, 
anticommute the heavy quark fields until they vanish on the vacuum state, compute the trace over the Pauli matrices, 
sum over the Clebsch--Gordan coefficients according to eq.~\eqref{eq:ClGo}, 
and finally sum over the angular wave functions according to eq. \eqref{eq:vl}.
We get
\begin{align}
    \bra{\Omega}\mathcal{O}^{P_{c\bar{c}}(4312)}({}^1S_0^{[8]})\ket{\Omega} =& \frac{a_1^2}{2\pi}\, |\phi_{(1/2)_g}(0)|^2 \,\mathcal{M}_{(1/2)_g}\,,
     \label{eq:LDMEPenta1S08}\\
    \bra{\Omega}\mathcal{O}^{P_{c\bar{c}}(4312)}({}^3S_1^{[8]})\ket{\Omega} =& \frac{a_3^2}{2\pi} \, \left|\phi_{(1/2)_g'}(0)\right|^2 \, \mathcal{M}_{(1/2)'_g} 
    + \frac{a_3^2}{2\pi} \, \left|\phi_{(3/2)_g}(0)\right|^2 \, \mathcal{M}_{(3/2)_g}\nonumber\\
    &+ \frac{a_2^2}{2\pi} \, \left|\phi_{(1/2)_g}(0)\right|^2 \, \mathcal{M}_{(1/2)_g}\,.
    \label{eq:LDMEPenta3S18}
\end{align}
Due to the orthogonality conditions in eqs.~\eqref{eq:ClGo} and \eqref{eq:vl}, we do not generate cross terms. 
The radial wave functions $\phi_{(1/2)_g}$, $\phi_{(1/2)_g'}$ and $\phi_{(3/2)_g}$ are solutions of the Schr\"odinger equations~\cite{Berwein:2024ztx,Brambilla:2025xma}
\begin{equation}
  \left[-\frac{1}{m_Qr^2}\,\partial_r\,r^2\,\partial_r
  +V_{(1/2)_g}(r)\right]\phi_{(1/2)_g}  =  E\,\phi_{(1/2)_g},
  \label{eq:schpenta1}
\end{equation}
and 
\begin{align}
  \Bigg[-\frac{1}{m_Qr^2}\,\partial_rr^2\partial_r+\frac{3}{m_Qr^2}
     \begin{pmatrix} 1 & -1 \\ -1 & 1\end{pmatrix} 
  +\begin{pmatrix} V_{(1/2)^\prime_g}(r) & 0 \\ 0 & V_{(3/2)_g}(r) \end{pmatrix}\Bigg] &
  \begin{pmatrix} \phi_{(1/2)'_g} \\ \phi_{(3/2)_g}\end{pmatrix}= \nonumber\\
      & \hspace{7mm}
  E \begin{pmatrix} \phi_{(1/2)'_g} \\ \phi_{(3/2)_g}\end{pmatrix},
   \label{eq:schpenta2}
\end{align}
with $V_{(1/2)_g}(r)$, $V_{(1/2)'_g}(r)$ and $V_{(3/2)_g}(r)$ the adiabatic BO potentials for pentaquarks  
in scenario I, where it is assumed that all three potentials support bound states.
The matrix elements are given by 
\begin{align}
     \mathcal{M}_{(1/2)_g} &=  \sum_X\sum_{\lambda = \pm 1/2}\left| \bra{\Omega}\Phi_{\ell}^{\dagger AB}T_{ji}^A\ket{X;(1/2)^+, (1/2)_g(\lambda); \bm{0} ;i,j} \right|^2 ,\\
    \mathcal{M}_{(1/2)_g'} &= \sum_X\sum_{\lambda = \pm 1/2}\left| \bra{\Omega}\Phi_{\ell}^{\dagger AB}T_{ji}^A\ket{X;(3/2)^+, (1/2)_g'(\lambda); \bm{0} ;i,j} \right|^2 ,\\
    \mathcal{M}_{(3/2)_g}  &= \sum_X\sum_{\lambda = \pm 3/2}\left| \bra{\Omega}\Phi_{\ell}^{\dagger AB}T_{ji}^A\ket{X; (3/2)^+, (3/2)_g(\lambda); \bm{0} ;i,j} \right|^2.
\end{align}

Through the unitary transformation 
\begin{equation}
  U=
\frac{1}{\sqrt{2}}\begin{pmatrix}
 1 & 1 \\
 1 & -1 \\
\end{pmatrix}
= U^{-1}
\label{eq:PentaTransformMatrix}
\end{equation}
we can diagonalize the centrifugal barrier in the coupled Schr\"odinger equations \eqref{eq:schpenta2}.
In the new (diabatic) basis, we identify an $S$-wave function $\phi_S$ and a $D$-wave function $\phi_D$.\footnote{
To reduce the proliferation of symbols, we use here the same symbols for the $S$-wave and $D$-wave radial 
wave functions and associated matrix elements as the ones used in the $\chi_{c1}(3872)$ case. 
It should be clear, however, that although the symbols are the same, wave functions for tetraquarks and pentaquarks are different as they satisfy different Schr\"odinger equations with different potentials.
Also, the matrix elements differ in the two cases as they refer to Fock states with different particle content and quantum numbers.}
The wave functions $\phi_S$ and $\phi_D$ follow from transforming according to $U$ the wave functions $\phi_{(1/2)_g}'$ and $\phi_{(3/2)_g}$,
\begin{equation}
U \begin{pmatrix}
 \phi_{(1/2)_g^\prime}\\
\phi_{(3/2)_g}  
\end{pmatrix} \equiv
\begin{pmatrix}
 \phi_{S}\\
 \phi_{D}
\end{pmatrix}.
\label{eq:Uphipenta}
\end{equation}
Solving the Schr\"odinger equations \eqref{eq:schpenta1} and \eqref{eq:schpenta2} as in~\cite{Brambilla:2025xma}, 
we obtain 
\begin{align}
    \left|\phi_{(1/2)_g}(0)\right|^2 &= 2.74\times 10^{-3} \, \mathrm{GeV}^3,
    \label{pentawave1}\\
    \left|\phi_S(0)\right|^2 &= 1.46 \times 10^{-2} \, \mathrm{GeV}^3, 
    \label{pentawave2}\\
    \left|\phi_D(0)\right|^2 &\approx 0 \, \mathrm{GeV}^3. 
    \label{pentawave3}
\end{align}
Because of the dominance of the $S$-wave and $(1/2)_g$ components by orders of magnitude, we can drop the $D$-wave component. 
By neglecting the $D$-wave component, the octet LDME \eqref{eq:LDMEPenta3S18} 
expressed in terms of the wave functions in the diabatic basis reads 
\begin{equation}
       \bra{\Omega}\mathcal{O}^{P_{c\bar{c}}(4312)}({}^3S_1^{[8]})\ket{\Omega} = 
       \frac{a_2^2}{2\pi}|\phi_{(1/2)_g}(0)|^2 \mathcal{M}_{(1/2)_g} 
       + \frac{a_3^2}{2\pi} \left|\phi_S(0)\right|^2\mathcal{M}_S\,,
       \label{eq:PentaLDME3S18_final}
\end{equation}
with $\mathcal{M}_S$ given by 
\begin{equation}
    \mathcal{M}_S = \frac{1}{2}\left( \mathcal{M}_{(1/2)_g'} + \mathcal{M}_{(3/2)_g} \right).
 \label{eq:UM}   
\end{equation}
The matrix element $\bra{\Omega}\mathcal{O}^{P_{c\bar{c}}(4312)}({}^1S_0^{[8]})\ket{\Omega}$ can be read off eq.~\eqref{eq:LDMEPenta1S08}.
Both LDMEs contribute to the unpolarized inclusive production cross section of the $P_{c\bar{c}}(4312)^+$
at leading order in $v$:
\begin{equation}
\sigma_{P_{c\bar{c}}(4312)^+} =   \sigma_{Q \bar Q({}^1S_0^{[8]})}\bra{\Omega} {\cal O}^{P_{c\bar{c}}(4312)} ({}^1S_0^{[8]}) \ket{\Omega} + \sigma_{Q \bar Q({}^3S_1^{[8]})}\bra{\Omega} {\cal O}^{P_{c\bar{c}}(4312)} ({}^3S_1^{[8]}) \ket{\Omega}.
\label{eq:NRQCDfacPcca}
\end{equation}
The short-distance coefficients $\sigma_{Q \bar Q({}^1S_0^{[8]})}$ and $\sigma_{Q \bar Q({}^3S_1^{[8]})}$ are computed at next-to-leading order in $\alpha_s$ as explained in section~\ref{sec:HadroXc}. 

The matrix elements $\bra{\Omega}\mathcal{O}^{P_{c\bar{c}}(4457)}({}^1S_0^{[8]})\ket{\Omega}$ and $\bra{\Omega}\mathcal{O}^{P_{c\bar{c}}(4457)}({}^3S_1^{[8]})\ket{\Omega}$
for the $P_{c\bar c}(44$ $57)^+$ state follow from eqs.~\eqref{eq:LDMEPenta1S08} 
and~\eqref{eq:PentaLDME3S18_final} by replacing $a_i$ with $b_i$.
They provide the low-energy part of the unpolarized inclusive production cross section of the $P_{c\bar{c}}(4457)^+$ at leading order in $v$:
\begin{equation}
\sigma_{P_{c\bar{c}}(4457)^+} =   \sigma_{Q \bar Q({}^1S_0^{[8]})}\bra{\Omega} {\cal O}^{P_{c\bar{c}}(4457)} ({}^1S_0^{[8]}) \ket{\Omega} + \sigma_{Q \bar Q({}^3S_1^{[8]})}\bra{\Omega} {\cal O}^{P_{c\bar{c}}(4457)} ({}^3S_1^{[8]}) \ket{\Omega}.
\label{eq:NRQCDfacPccb}
\end{equation}

Along the same lines, we can compute the octet LDMEs for the production of the $P_{c\bar c}(4380)^+$ and $P_{c\bar c}(4440)^+$ states.
The only difference is that, since these two states are assumed to be a $J=3/2$ state, 
the relevant states entering the matrix element are \eqref{eq:BOEFTPenta12a} and \eqref{eq:BOEFTPenta32}.
Neglecting again the $D$-wave component, we obtain
\begin{align}
    \langle\mathcal{O}^{P_{c\bar c}(4380)}({}^1S_0^{[8]})\rangle &= 
    \frac{c_2^2}{\pi}\left| \phi_S(0) \right|^2 \mathcal{M}_S \,,
    \label{eq:PentaLDME1S08_4380}\\
    \langle\mathcal{O}^{P_{c\bar c}(4380)}({}^3S_1^{[8]})\rangle &= 
    \frac{c_1^2}{\pi}\left| \phi_{(1/2)_g}(0) \right|^2 \mathcal{M}_{(1/2)_g} 
    + \frac{c_3^2}{\pi} \left| \phi_S(0) \right|^2 \mathcal{M}_S\,,
   \label{eq:PentaLDME3S18_4380} 
\end{align}
which, inserted in
\begin{equation}
\sigma_{P_{c\bar{c}}(4380)^+} =   \sigma_{Q \bar Q({}^1S_0^{[8]})}\bra{\Omega} {\cal O}^{P_{c\bar{c}}(4380)} ({}^1S_0^{[8]}) \ket{\Omega} + \sigma_{Q \bar Q({}^3S_1^{[8]})}\bra{\Omega} {\cal O}^{P_{c\bar{c}}(4380)} ({}^3S_1^{[8]}) \ket{\Omega}, 
\label{eq:NRQCDfacPccc}
\end{equation}
provides the unpolarized inclusive production cross section of the $P_{c\bar{c}}(4380)^+$ at leading order in $v$.

The matrix elements  $\bra{\Omega}\mathcal{O}^{P_{c\bar{c}}(4440)}({}^1S_0^{[8]})\ket{\Omega}$ and $\bra{\Omega}\mathcal{O}^{P_{c\bar{c}}(4440)}({}^3S_1^{[8]})\ket{\Omega}$
for the $P_{c \bar c}(44$ $40)^+$ state follow from eqs.~\eqref{eq:PentaLDME1S08_4380} and \eqref{eq:PentaLDME3S18_4380} by replacing $c_i$ with $d_i$.
They provide the low-energy part of the unpolarized inclusive production cross section of the $P_{c\bar{c}}(4440)^+$ at leading order in $v$:
\begin{equation}
\sigma_{P_{c\bar{c}}(4440)^+} =   \sigma_{Q \bar Q({}^1S_0^{[8]})}\bra{\Omega} {\cal O}^{P_{c\bar{c}}(4440)} ({}^1S_0^{[8]}) \ket{\Omega} + \sigma_{Q \bar Q({}^3S_1^{[8]})}\bra{\Omega} {\cal O}^{P_{c\bar{c}}(4440)} ({}^3S_1^{[8]}) \ket{\Omega}.
\label{eq:NRQCDfacPccd}
\end{equation}

As in the case of the $\chi_{c1}(3872)$ production, by completing the sum of states in $\mathcal{M}_{(1/2)_g}$ and $\mathcal{M}_S$ we can put some upper limits on $\mathcal{M}_{(1/2)_g}$ and $\mathcal{M}_S$. 
The upper limits read $\mathcal{M}_{(1/2)_g} \le (1/2)(N_c^2-1) = 4$ and $\mathcal{M}_S \le (1/2)(N_c^2-1)/2 = 2$.

A further constraint comes from the pentaquark production in $b$-hadron decays:
\begin{align}
    \mathrm{Br}(\Lambda_b \to P_{c\bar{c}}^+ +X) &= \mathrm{Br}(b \rightarrow \bar{c}c({}^3S_1^{[8]})+X)\bra{\Omega}\mathcal{O}^{P_{c\bar{c}}}({}^3S_1^{[8]})\ket{\Omega} \nonumber\\
    &\quad + \mathrm{Br}(b \rightarrow \bar{c}c({}^1S_0^{[8]})+X)\bra{\Omega}\mathcal{O}^{P_{c\bar{c}}}({}^1S_0^{[8]})\ket{\Omega}. \label{eq:BranchingPenta}
\end{align}
The right-hand side depends on the two branching ratios 
$\mathrm{Br}(b \rightarrow \bar{c}c({}^3S_1^{[8]})+X)$ and $\mathrm{Br}(b \rightarrow \bar{c}c({}^1S_0^{[8]})+X)$. 
From the expressions derived in~\cite{Beneke:1998ks}, we obtain, again with masses $m_c = 1.4 \, \mathrm{GeV}$ and $m_b = 4.74 \, \mathrm{GeV}$ and varying the renormalization scale between $\mu = m_b/2$ and $\mu  = 2m_b$, that 
${\mathrm{Br}(b\to c\bar{c}({}^1S_0^{[8]})+X)} = 0.395^{+0.034}_{-0.049} \, \mathrm{GeV}^{-3}$ and 
 $\mathrm{Br}(b\to c\bar{c}({}^3S_1^{[8]})$ $+X) = 0.213^{+0.019}_{-0.028} \, \mathrm{GeV}^{-3}$. The latter value has already been derived in  section~\ref{sec:X3872Bdecay}.
For what concerns the left-hand side of eq. \eqref{eq:BranchingPenta}, 
experimentally there are only measurements for $\mathrm{Br}(\Lambda_b \to P_{c\bar{c}}^+K^-)\times \mathrm{Br}(P_{c\bar{c}}^+\to p J/\psi )$ available, and also only for the case of $P_{c\bar c}(4457)^+$ and $P_{c\bar c}(4380)^+$, namely~\cite{LHCb:2015oyu}  
\begin{align}
 \mathrm{Br}(\Lambda_b \to P_{c\bar{c}}(4457)^+K^-)\times \mathrm{Br}(P_{c\bar{c}}(4457)^+\to p J/\psi ) &= (1.3 \pm 0.4)\times 10^{-5}, 
 \label{BrPcc4457exp}\\
\mathrm{Br}(\Lambda_b \to P_{c\bar{c}}(4380)^+K^-)\times \mathrm{Br}(P_{c\bar{c}}(4380)^+\to p J/\psi ) &= (2.7 \pm 1.4)\times 10^{-5}.
\label{BrPcc4380exp}
\end{align}
For $P_{c\bar c}(4312)^+$, only an upper bound for its decay rate into~$\eta_c$ has been measured. The exclusive branching ratios~\eqref{BrPcc4457exp} and~\eqref{BrPcc4380exp} are, however, sufficient to estimate  
lower limits for $\mathrm{Br}(\Lambda_b \to P_{c\bar{c}}(4457)^+ +X)$ and $\mathrm{Br}(\Lambda_b \to P_{c\bar{c}}(4380)^+ +X)$, due to the relation
\begin{equation}
    \mathrm{Br}(\Lambda_b \to P_{c\bar{c}}^+ +X) 
    >\frac{ \mathrm{Br}(\Lambda_b \to P_{c\bar{c}}^+K^-)\times \mathrm{Br}(P_{c\bar{c}}^+\to p J/\psi )}{\mathrm{Br}(P_{c\bar c}^+\to J/\psi +X)} ,
\label{eq:lowerBr}
\end{equation}
where we can calculate the denominator by combining the partial decay width, determined in~\cite{Brambilla:2025xma}, with total decay rates reported in~\cite{LHCb:2015yax}, resulting in
\begin{align}
\mathrm{Br}(P_{c\bar{c}}(4457)^+ \to J/\psi + X) &= 0.172^{+0.186}_{-0.087}  \,,   \label{eq:Penta4457_Brcalc}\\ 
\mathrm{Br}(P_{c\bar{c}}(4380)^+ \to J/\psi + X) &= 0.085^{+0.077}_{-0.041} \,.
\label{eq:Penta4380_Brcalc}
\end{align}
Here, we have used Monte Carlo simulations to propagate the large, asymmetric uncertainties of the partial widths and total decay rates following the method described in ref.~\cite{Crowder2020}:
For a given $x^{+\sigma_+}_{-\sigma_-}$, we assume that the distribution is a superposition of two Gaussians with variances $\sigma_+$ and $\sigma_-$ respectively, resulting in a skewed Gaussian distribution. 
The central values in \eqref{eq:Penta4457_Brcalc} and \eqref{eq:Penta4380_Brcalc} are the medians of the obtained distributions, and the upper and lower bounds represent the $1\sigma$ interval.
The lower limits \eqref{eq:lowerBr} for the $P_{c\bar{c}}(4457)^+$ and the $P_{c\bar{c}}(4380)^+$ 
applied to the left-hand side of eq. \eqref{eq:BranchingPenta} lead to lower limits 
on the matrix elements $\mathcal{M}_S$ and $\mathcal{M}_{(1/2)_g}$, due to the positivity 
of all coefficients in eqs. \eqref{eq:PentaLDME1S08_4380}-\eqref{eq:NRQCDfacPccc}.
We calculate the total uncertainties of the lower limits of $\mathcal{M}_S$ and $\mathcal{M}_{(1/2)_g}$ by combining the scale uncertainties of the short-distance coefficients, estimated as before, the experimental uncertainties in \eqref{BrPcc4457exp} and \eqref{BrPcc4380exp}, 
the uncertainties in the theoretical calculation of the branching ratios \eqref{eq:Penta4457_Brcalc} and \eqref{eq:Penta4380_Brcalc}, and a 30\% uncertainty due to neglected $\mathcal{O}(v^2)$ terms in the right-hand side of eq.~\eqref{eq:BranchingPenta}, again performing the error propagation by means of the Monte Carlo simulation method~\cite{Crowder2020}.

\begin{figure}[ht]
    \centering
    \begin{subfigure}{0.48\textwidth}
        \includegraphics[scale = 0.39]{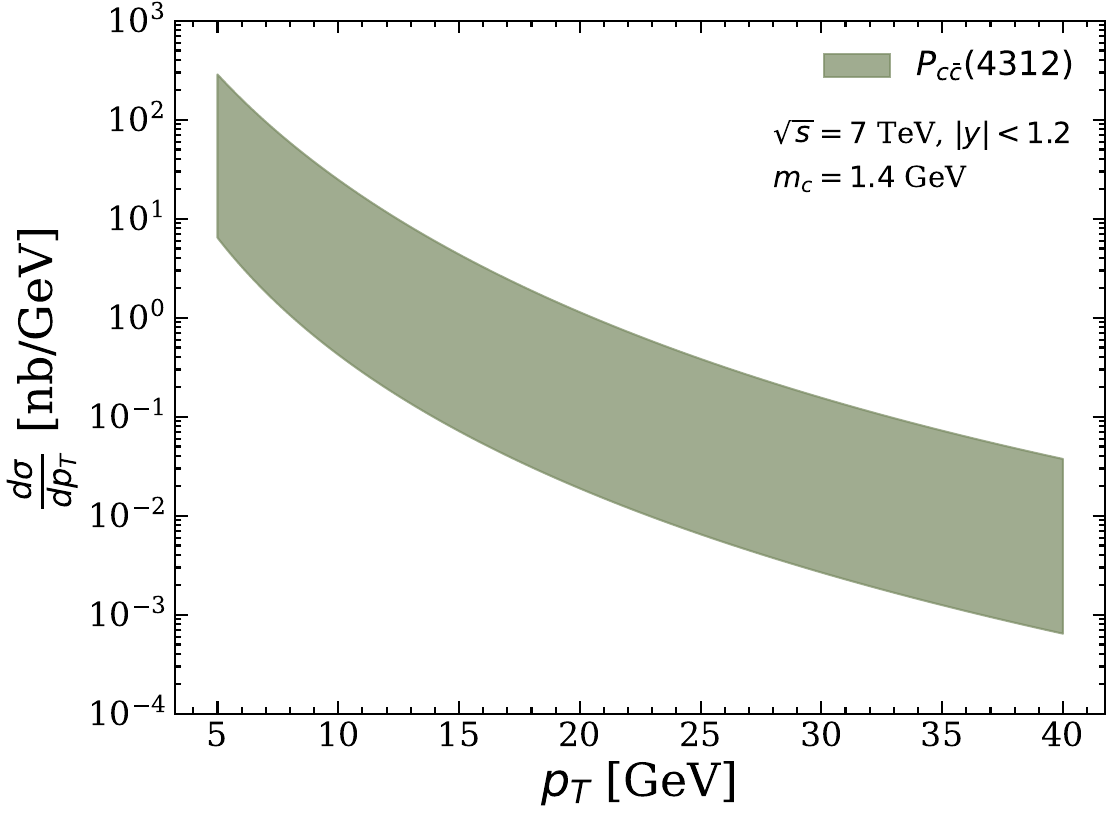}
    \end{subfigure}
    \begin{subfigure}{0.48\textwidth}
        \includegraphics[scale = 0.39]{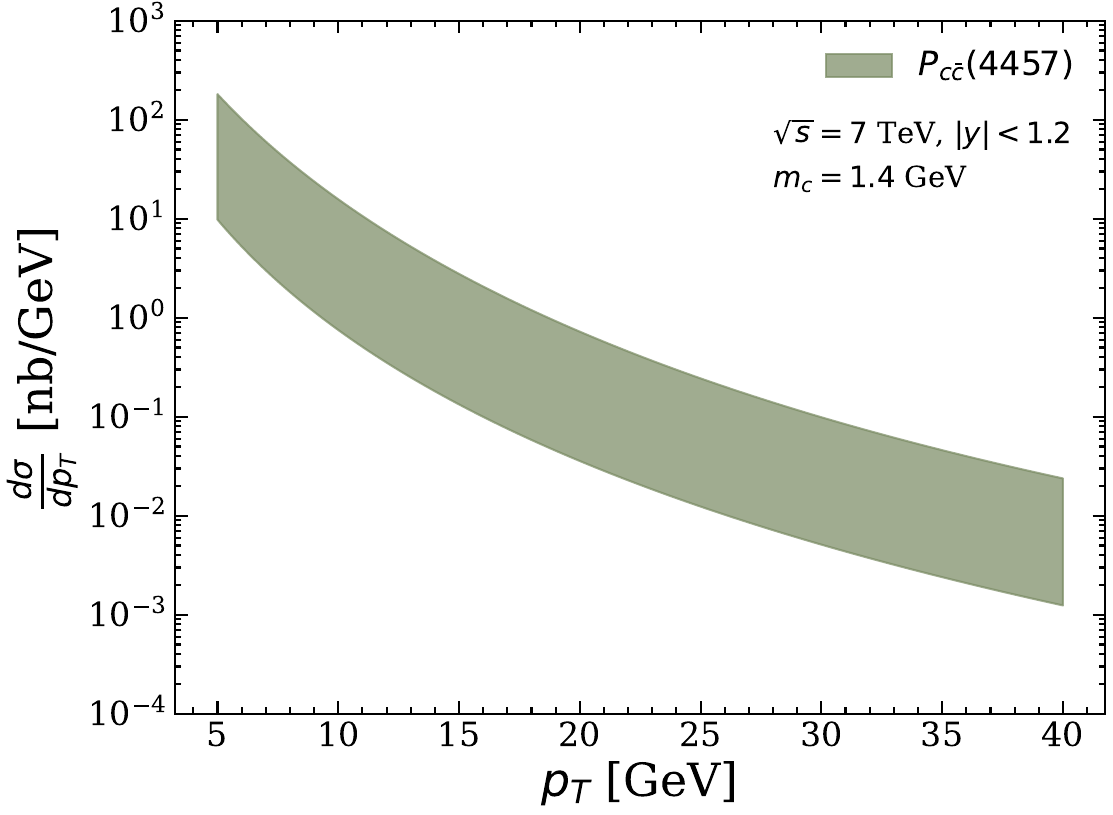}
    \end{subfigure}
    \\
    \begin{subfigure}{0.48\textwidth}
        \includegraphics[scale = 0.39]{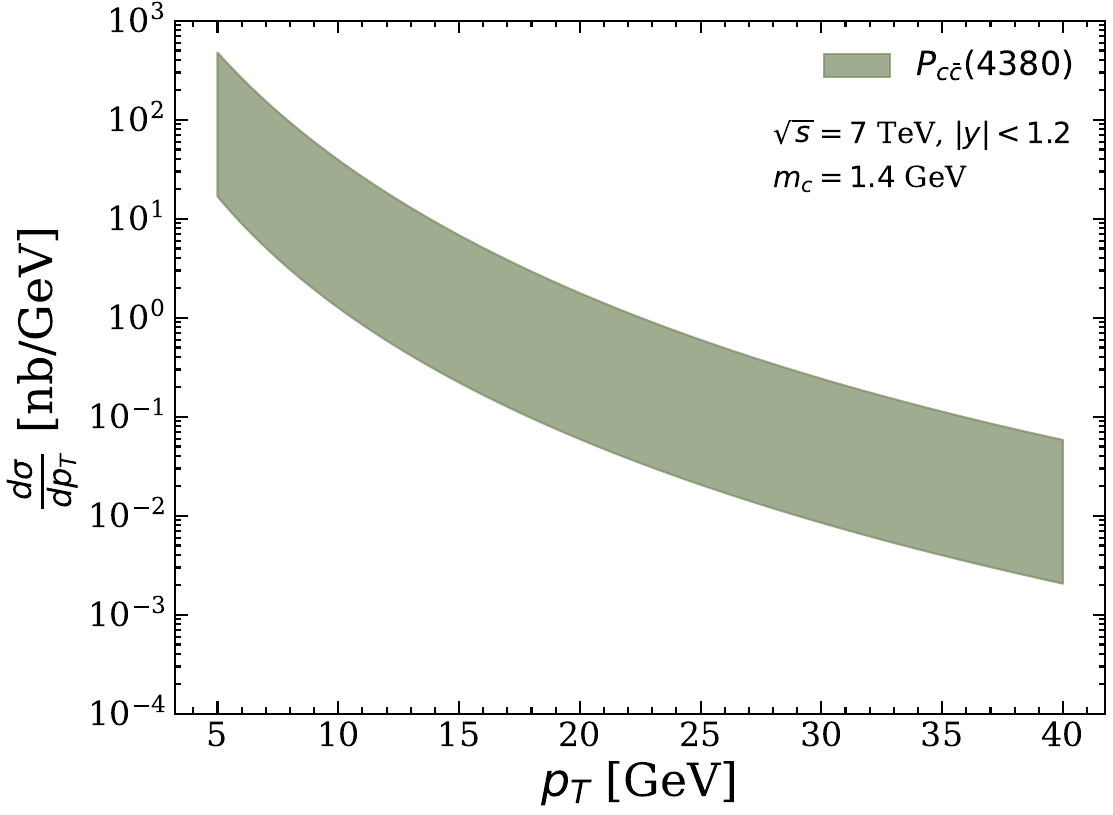}
    \end{subfigure}
    \begin{subfigure}{0.48\textwidth}
        \includegraphics[scale = 0.39]{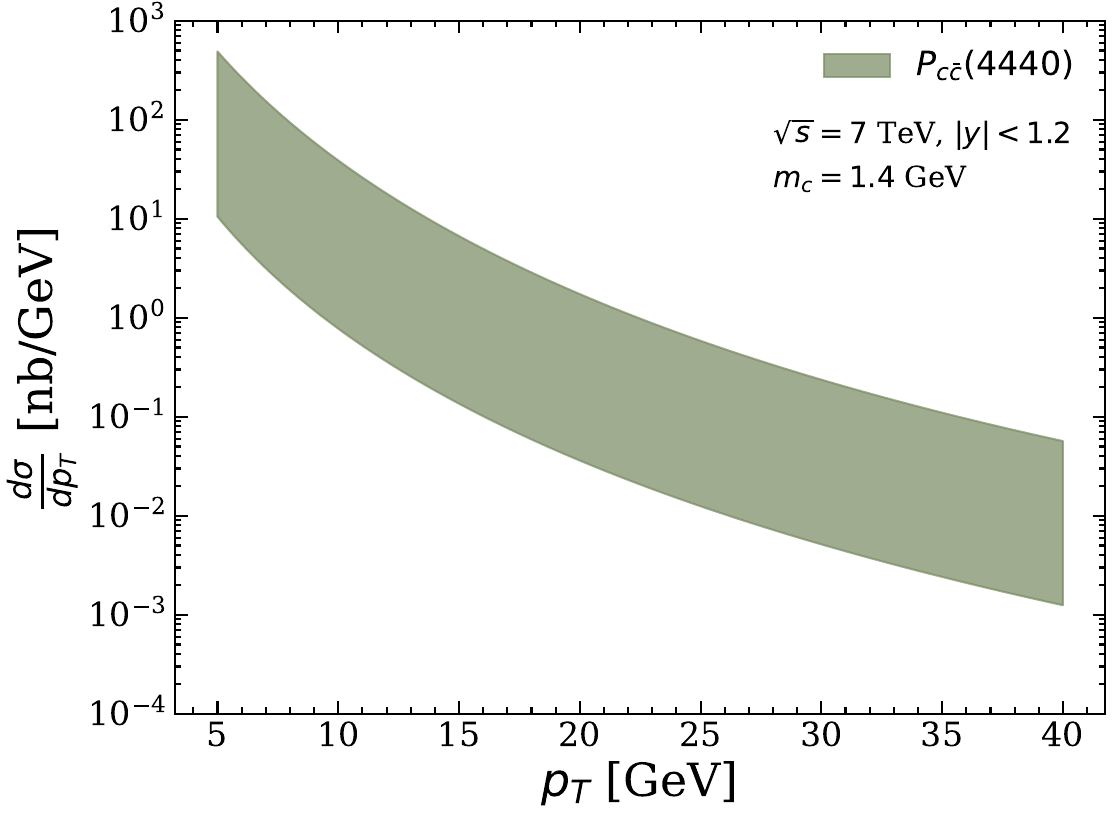}
    \end{subfigure}
    \caption{Predicted prompt inclusive differential hadroproduction cross sections of the
    charmonium pentaquark states 
    $P_{c\bar{c}}(4312)^+$,  $P_{c\bar{c}}(4457)^+$,  $P_{c\bar{c}}(4380)^+$ and $P_{c\bar{c}}(4440)^+$ within scenario I.}
    \label{fig:PentaCharmScI}
\end{figure}

Combined with the upper limits from BOEFT, $\mathcal{M}_{(1/2)_g} \le 4$ and $\mathcal{M}_S \le 2$,
and the requirement of, in absence of renormalization effects, positive LDMEs, we thus obtain
\begin{align}
0.024^{+ 0.14}_{-0.024} \lesssim  &\, \mathcal{M}_S \le 2 \,, \label{eq:NumericalMIa}\\
0.608^{+0.958}_{-0.414}\lesssim  &\, \mathcal{M}_{(1/2)_g}  \le  4\,. \label{eq:NumericalMIb}
\end{align}
For the individual LDMEs of the various pentaquark states, these limits translate into
\begin{align}
9.50^{+8.28}_{-6.62} \times 10^{-5} \lesssim\, &\bra{\Omega}\mathcal{O}^{P_{c\bar{c}}(4312)}({}^1S_0^{[8]})\ket{\Omega} \lesssim 3.94 \times 10^{-4}
\,\mathrm{GeV}^3,
\label{eq:Pcc4312uppera}\\
1.94^{+1.82}_{-1.34} \times 10^{-4} \lesssim\, &\bra{\Omega}\mathcal{O}^{P_{c\bar{c}}(4312)}({}^3S_1^{[8]})\ket{\Omega} \lesssim 3.38 \times 10^{-3} \,\mathrm{GeV}^3,
\label{eq:Pcc4312upperb}\\
4.61^{+4.03}_{-3.21} \times 10^{-5} \lesssim\, &\bra{\Omega}\mathcal{O}^{P_{c\bar{c}}(4457)}({}^1S_0^{[8]})\ket{\Omega} \lesssim  1.91 \times 10^{-4} \,\mathrm{GeV}^3,
\label{eq:Pcc4457uppera}\\
3.43^{+2.62}_{-2.13} \times 10^{-4} \lesssim\, &\bra{\Omega}\mathcal{O}^{P_{c\bar{c}}(4457)}({}^3S_1^{[8]})\ket{\Omega} \lesssim 2.15 \times 10^{-3} \,\mathrm{GeV}^3,
\label{eq:Pcc4457upperb}\\
1.17^{+1.28}_{-0.94} \times 10^{-4} \lesssim\, &\bra{\Omega}\mathcal{O}^{P_{c\bar{c}}(4380)}({}^1S_0^{[8]})\ket{\Omega} \lesssim 2.38 \times 10^{-3} \,\mathrm{GeV}^3,
\label{eq:Pcc4380uppera}\\
4.79^{+2.35}_{-2.75} \times 10^{-4} \lesssim\, &\bra{\Omega}\mathcal{O}^{P_{c\bar{c}}(4380)}({}^3S_1^{[8]})\ket{\Omega} \lesssim 5.25 \times 10^{-3} \,\mathrm{GeV}^3,
\label{eq:Pcc4380upperb}\\
1.78^{+1.96}_{-1.43} \times 10^{-4} \lesssim\, &\bra{\Omega}\mathcal{O}^{P_{c\bar{c}}(4440)}({}^1S_0^{[8]})\ket{\Omega} \lesssim 3.62 \times 10^{-3} \,\mathrm{GeV}^3,
\label{eq:Pcc4440uppera}\\
3.27^{+2.74}_{-2.09} \times 10^{-4} \lesssim\, &\bra{\Omega}\mathcal{O}^{P_{c\bar{c}}(4440)}({}^3S_1^{[8]})\ket{\Omega} \lesssim 5.11 \times 10^{-3} \,\mathrm{GeV}^3.
\label{eq:Pcc4440upperb}
\end{align}

Finally, we display the results for the differential pentaquark production cross sections in figure~\ref{fig:PentaCharmScI}. Thereby, we determine the uncertainty bands using Monte Carlo samples propagating the errors of $\mathcal{M}_{(1/2)_g}$ and $\mathcal{M}_S$ according to Eqs.~\eqref{eq:NumericalMIa} and~\eqref{eq:NumericalMIb}, using the respective lower limits, and the scale uncertainties of the short distance cross sections as in section \ref{sec:HadroXc}.
In all four cases, the cross sections of the pentaquark states are comparable in magnitude and shape to the $\chi_{c1}(3872)$ one, as shown in the first plot of figure~\ref{fig:SigmaXc}, but with much larger uncertainties.

\subsection{Hadroproduction of charmonium pentaquarks in scenario II}
In scenario II, the pentaquark states $P_{c\bar{c}}(4312)^+$, $P_{c\bar{c}}(4457)^+$, $P_{c\bar{c}}(4380)^+$ 
and $P_{c\bar{c}}(4440)^+$ are written as~\cite{Alasiri:2025roh} 
\begin{align}  
\ket{P_{c\bar c}(4312)^+} &= \ket{S=1;k = 3/2}_{J^P=(1/2)^-},
\label{eq:Penta4312stateABB}\\
\ket{P_{c\bar c}(4457)^+} &= \ket{S=1; k=3/2}_{J^P=(5/2)^-},
\label{eq:Penta4457stateABB}\\
\ket{P_{c\bar c}(4380)^+} &= d_1 \ket{S=0;k = 3/2}_{J^P = (3/2)^-} + d_2 \ket{S=1; k=3/2}_{J^P = (3/2)^-},
\label{eq:Penta4380stateABB} \\
\ket{P_{c\bar c}(4440)^+} &= -d_2 \ket{S=0;k = 3/2}_{J^P = (3/2)^-} + d_1 \ket{S=1; k=3/2}_{J^P = (3/2)^-},
\label{eq:Penta4440stateABB} 
\end{align}
with $d_1 = -0.368$ and $d_2 = 0.930$. 
The eigenstates of the heavy-quark-antiquark spin and the total angular momentum of the light quarks, 
$\ket{S;k}_{J^P}$, can be read off eqs.~\eqref{eq:BOEFTPenta12}-\eqref{eq:BOEFTPenta32} just by adpting 
the total angular momentum label, $J$, in the Clebsch--Gordan coefficients.

We start considering the pentaquark state $\ket{P_{c\bar c}(4312)^+}$ given in eq.~\eqref{eq:Penta4312stateABB}.
Compared to eq.~\eqref{eq:PentaState4312}, in this scenario only one state with $k^P=(3/2)^+$ and $S=1$ contributes. 
The state is given in eq.~\eqref{eq:BOEFTPenta32a}.
Proceeding like in scenario I, we obtain the BOEFT factorization formula
\begin{align}
    \bra{\Omega}\mathcal{O}^{P_{c\bar{c}}(4312)}({}^3S_1^{[8]})\ket{\Omega} &= 
    \frac{1}{2\pi} \left|\phi_{S}(0)\right|^2\mathcal{M}_{S}\,.
    \label{eq:LDME3S18PentaScenarioABB}
\end{align}
The radial wave function $\phi_S$ and the matrix element $\mathcal{M}_{S}$ are defined as in scenario I, 
see eqs. \eqref{eq:Uphipenta} and \eqref{eq:UM}, respectively.
The value of the wave function at the origin squared in this scenario is given by
\begin{align}
    |\phi_S(0)|^2 &= 3.04\times 10^{-2} \, \mathrm{GeV}^3.
\label{eq:wf-scenarioII}    
\end{align}
The LDME \eqref{eq:LDME3S18PentaScenarioABB} encodes the complete low-energy contribution to the 
unpolarized inclusive production cross section of the $P_{c\bar{c}}(4312)^+$ at leading order in $v$:
\begin{equation}
\sigma_{P_{c\bar{c}}(4312)^+} =  \sigma_{Q \bar Q({}^3S_1^{[8]})}\bra{\Omega} {\cal O}^{P_{c\bar{c}}(4312)} ({}^3S_1^{[8]}) \ket{\Omega}.
\label{eq:NRQCDfacPcc4312II}
\end{equation}
This is a consequence of the state $\ket{P_{c\bar c}(4312)^+}$ being in scenario II an eigenstate of the heavy quark spin with $S=1$.

The state $\ket{P_{c\bar c}(4457)^+}$ is, like the state $\ket{P_{c\bar c}(4312)^+}$, a state with definite heavy quark spin, see eq. \eqref{eq:Penta4457stateABB}.
Contrary to the other three states, this state has different $J^P$ quantum numbers in the two scenarios:
in scenario I it has $J^P = (1/2)^-$, while in the scenario II it has $J^P=(5/2)^-$. 
In the BOEFT, it holds the factorization formula  
\begin{align}
    \bra{\Omega}\mathcal{O}^{P_{c\bar{c}}(4457)}({}^3S_1^{[8]})\ket{\Omega} &= 
    \frac{3}{2\pi} \left|\phi_S(0)\right|^2\mathcal{M}_S\,.
\label{eq:LDME3S18PentaScenarioABBb}  
\end{align}
The factor 3 difference with respect to \eqref{eq:LDME3S18PentaScenarioABB} originates from the difference in $J$ when computing the sum over the Clebsch--Gordan coefficients \eqref{eq:ClGo}.
Again, because the state $\ket{P_{c\bar c}(4457)^+}$ in scenario II has definite heavy quark spin $S=1$, 
the octet LDME \eqref{eq:LDME3S18PentaScenarioABBb} encodes the complete low-energy contribution to the 
unpolarized inclusive production cross section of the $P_{c\bar{c}}(4457)^+$ at leading order in $v$:
\begin{equation}
\sigma_{P_{c\bar{c}}(4457)^+} =  \sigma_{Q \bar Q({}^3S_1^{[8]})}\bra{\Omega} {\cal O}^{P_{c\bar{c}}(4457)} ({}^3S_1^{[8]}) \ket{\Omega}.
\label{eq:NRQCDfacPcc4457II}
\end{equation}

The states $\ket{P_{c\bar c}(4380)^+}$ and $\ket{P_{c\bar c}(4440)^+}$ are the only states in scenario II that are superpositions of heavy quark spin one and spin zero states, 
see eqs.~\eqref{eq:Penta4380stateABB} and \eqref{eq:Penta4440stateABB}.
Therefore, both octet LDMEs, $\bra{\Omega}\mathcal{O}^{P_{c\bar c}(4380)}({}^1S_0^{[8]})\ket{\Omega}$ 
and $\bra{\Omega}\mathcal{O}^{P_{c\bar c}(4380)}({}^3S_1^{[8]})\ket{\Omega}$ 
for the $P_{c\bar c}(4380)^+$, $\bra{\Omega}\mathcal{O}^{P_{c\bar c}(4440)}({}^1S_0^{[8]})\ket{\Omega}$ and  
$\bra{\Omega}\mathcal{O}^{P_{c\bar c}(4440)}({}^3S_1^{[8]})\ket{\Omega}$ 
for the $P_{c\bar c}(4440)^+$, contribute to the low-energy dynamics of the unpolarized inclusive production cross section at leading order in $v$, which reads like in eq.~\eqref{eq:NRQCDfacPccc} and~\eqref{eq:NRQCDfacPccd}, respectively.
The octet LDMEs factorize in the BOEFT as 
\begin{align}
        \bra{\Omega}\mathcal{O}^{P_{c\bar c}(4380)}({}^1S_0^{[8]})\ket{\Omega} &= 
        \frac{d_1^2}{\pi}\left| \phi_S(0) \right|^2 \mathcal{M}_S \,,
        \label{eq:LDME1S08PentaScenarioABB}\\
    \bra{\Omega}\mathcal{O}^{P_{c\bar c}(4380)}({}^3S_1^{[8]})\ket{\Omega} &= 
    \frac{d_2^2}{\pi} \left| \phi_S(0) \right|^2 \mathcal{M}_S\,,
    \label{eq:LDME3S18PentaScenarioABBc}\\
   \bra{\Omega}\mathcal{O}^{P_{c\bar c}(4440)}({}^1S_0^{[8]})\ket{\Omega} & = 
   \frac{d_2^2}{\pi}\left| \phi_S(0) \right|^2 \mathcal{M}_S \,,
   \label{eq:LDME1S08PentaScenarioABBe}\\ 
   \bra{\Omega}\mathcal{O}^{P_{c\bar c}(4440)}({}^3S_1^{[8]})\ket{\Omega} & = 
   \frac{d_1^2}{\pi} \left| \phi_S(0) \right|^2 \mathcal{M}_S\,.
    \label{eq:LDME3S18PentaScenarioABBf}
\end{align}

\begin{figure}[ht]
    \centering
    \begin{subfigure}{0.48\textwidth}
        \includegraphics[scale = 0.39]{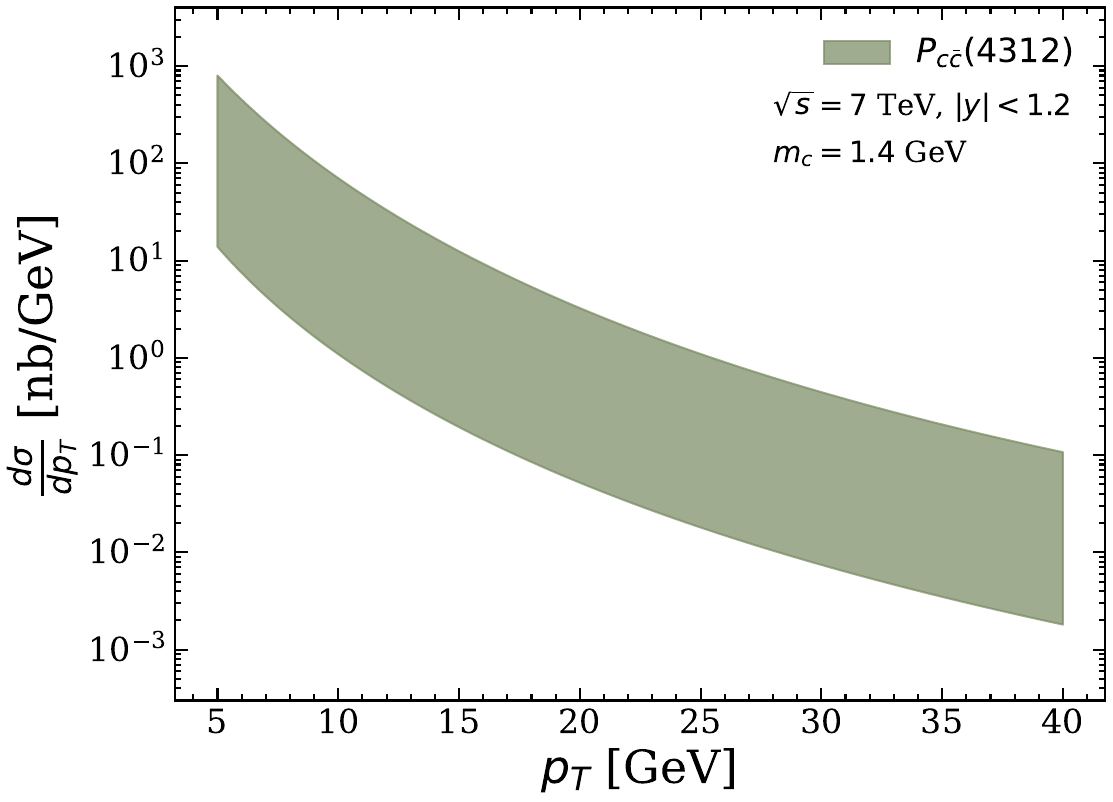}
    \end{subfigure}
    \begin{subfigure}{0.48\textwidth}
        \includegraphics[scale = 0.39]{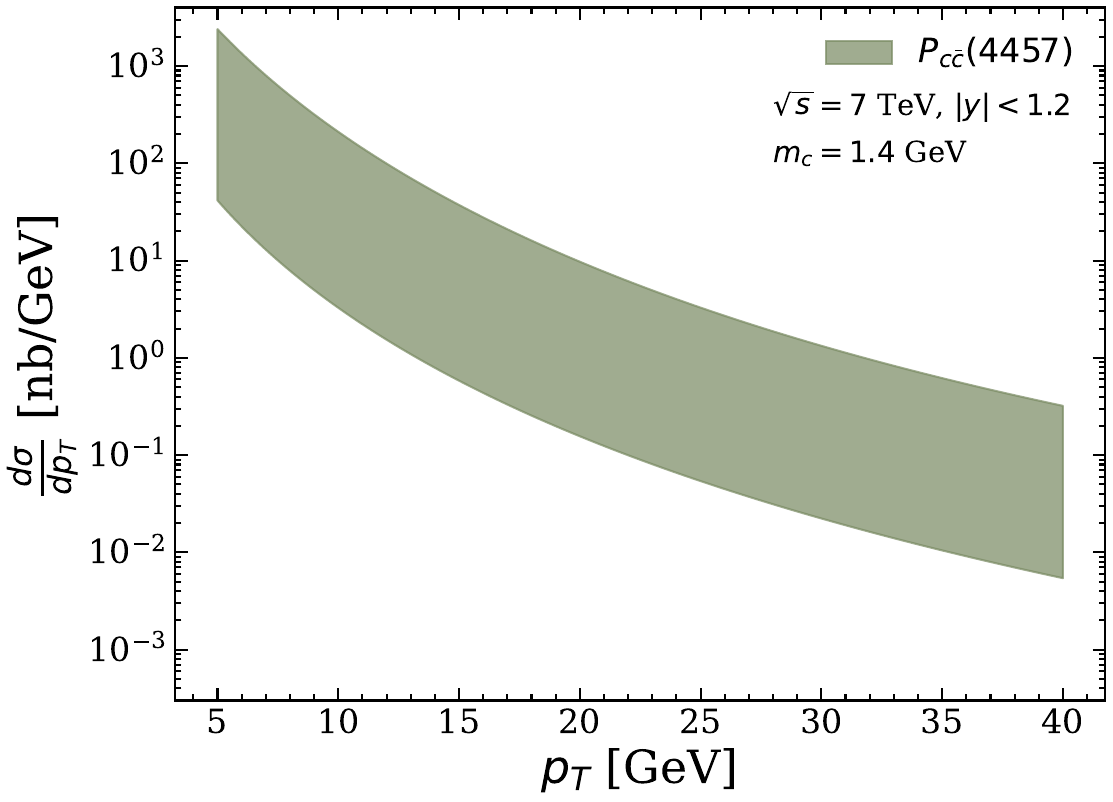}
    \end{subfigure}
    \\
    \begin{subfigure}{0.48\textwidth}
        \includegraphics[scale = 0.39]{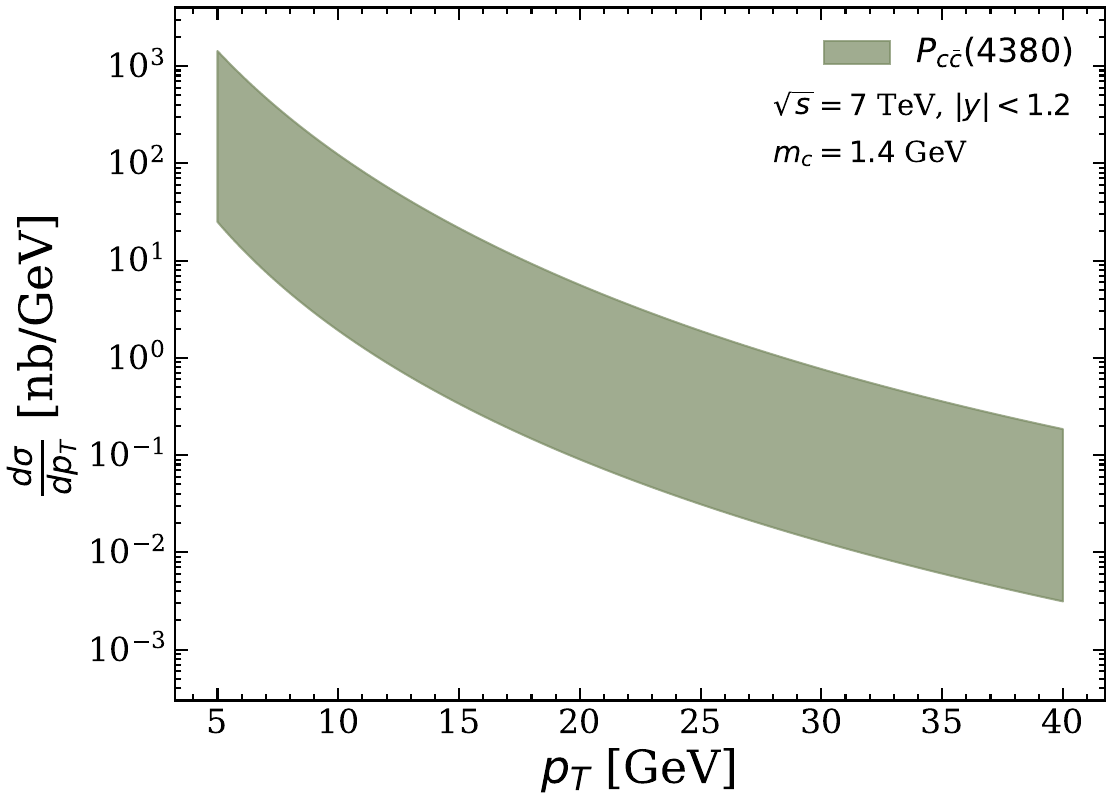}
    \end{subfigure}
    \begin{subfigure}{0.48\textwidth}
        \includegraphics[scale = 0.39]{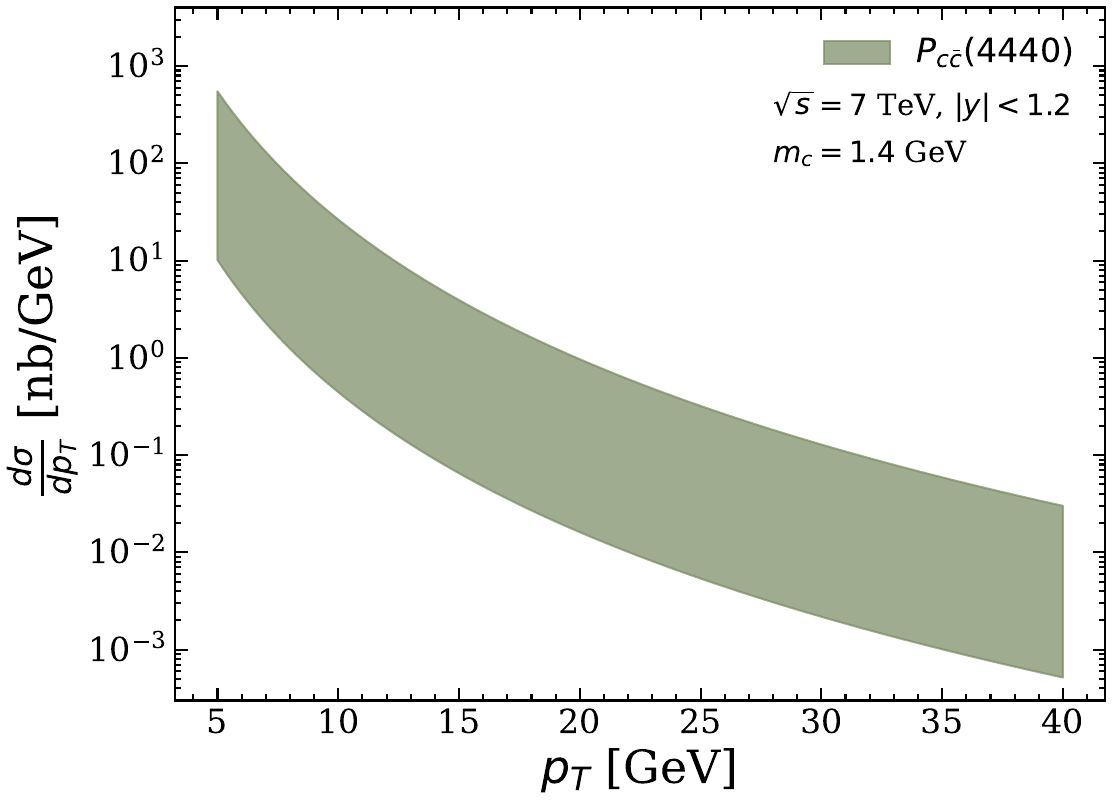}
    \end{subfigure}
    \caption{Predicted prompt inclusive differential hadroproduction cross sections of the
    charmonium pentaquark states 
    $P_{c\bar{c}}(4312)^+$,  $P_{c\bar{c}}(4457)^+$,  $P_{c\bar{c}}(4380)^+$ and $P_{c\bar{c}}(4440)^+$ within scenario II.}
    \label{fig:PentaCharmScII}
\end{figure}

To make predictions also in this scenario, we combine the upper limit on $\mathcal{M}_S$ dictated by the BOEFT with the lower limit that can be extracted from \eqref{BrPcc4380exp} following the same procedure as in scenario I; we obtain\footnote{
If we had fixed the lower limit using the branching fraction combination \eqref{BrPcc4457exp} for the state $P_{c\bar c}(4457)^+$, we would have obtained $\mathcal{M}_S \gtrsim 0.0181^{+0.0241}_{-0.0107}$. 
Both lower limits are consistent within the given uncertainties, 
but the one from the $P_{c\bar c}(4380)^+$ decay gives a stronger constraint on the matrix element.} 
\begin{align}
0.107^{+0.144}_{-0.072} \lesssim  \mathcal{M}_S \lesssim 2\,.
\label{eq:NumericalMII}
\end{align}
Using the BOEFT factorizations of the octet LDMEs given in~\eqref{eq:LDME3S18PentaScenarioABB}, \eqref{eq:LDME3S18PentaScenarioABBb}, and \eqref{eq:LDME1S08PentaScenarioABB}-\eqref{eq:LDME3S18PentaScenarioABBf},
and the wave function at the origin squared given in~\eqref{eq:wf-scenarioII}, the lower and upper bounds on $\mathcal{M}_S$ translate into the following lower and upper bounds on the octet LDMEs in scenario II:
\begin{align}
& 5.18^{+6.78}_{-3.13} \times 10^{-4} \lesssim \bra{\Omega}\mathcal{O}^{P_{c\bar{c}}(4312)}({}^3S_1^{[8]})\ket{\Omega} 
 \lesssim 9.68 \times 10^{-3} \,\mathrm{GeV}^3,\\
& 1.55^{+2.04}_{-0.94} \times 10^{-3} \lesssim \bra{\Omega}\mathcal{O}^{P_{c\bar{c}}(4457)}({}^3S_1^{[8]})\ket{\Omega} 
 \lesssim 2.90 \times 10^{-2} \,\mathrm{GeV}^3,\\
& 1.40^{+1.84}_{-0.85} \times 10^{-4} \lesssim \bra{\Omega}\mathcal{O}^{P_{c\bar{c}}(4380)}({}^1S_0^{[8]})\ket{\Omega}  
\lesssim 2.62 \times 10^{-3} \,\mathrm{GeV}^3,\\
&  8.96^{+11.7}_{-5.42} \times 10^{-4} \lesssim \bra{\Omega}\mathcal{O}^{P_{c\bar{c}}(4380)}({}^3S_1^{[8]})\ket{\Omega} 
\lesssim 1.67 \times 10^{-2} \,\mathrm{GeV}^3,\\
& 8.96^{+11.7}_{-5.42} \times 10^{-4} \lesssim \bra{\Omega}\mathcal{O}^{P_{c\bar{c}}(4440)}({}^1S_0^{[8]})\ket{\Omega}  
\lesssim 1.67 \times 10^{-2}  \,\mathrm{GeV}^3,\\
& 1.40^{+1.84}_{-0.85} \times 10^{-4} \lesssim \bra{\Omega}\mathcal{O}^{P_{c\bar{c}}(4440)}({}^3S_1^{[8]})\ket{\Omega} 
\lesssim  2.62 \times 10^{-3} \,\mathrm{GeV}^3.
\end{align}

Finally, we show in this scenario the predictions for the differential cross sections in figure~\ref{fig:PentaCharmScII}.
The uncertainties are estimated as in scenario I.
In all four cases and at the present level of accuracy, the results for scenarios I and II largely overlap within the error bands.

\subsection{Hadroproduction of bottomonium pentaquarks}
\begin{figure}[ht]
    \centering
    \begin{subfigure}{0.48\textwidth}
        \includegraphics[scale = 0.39]{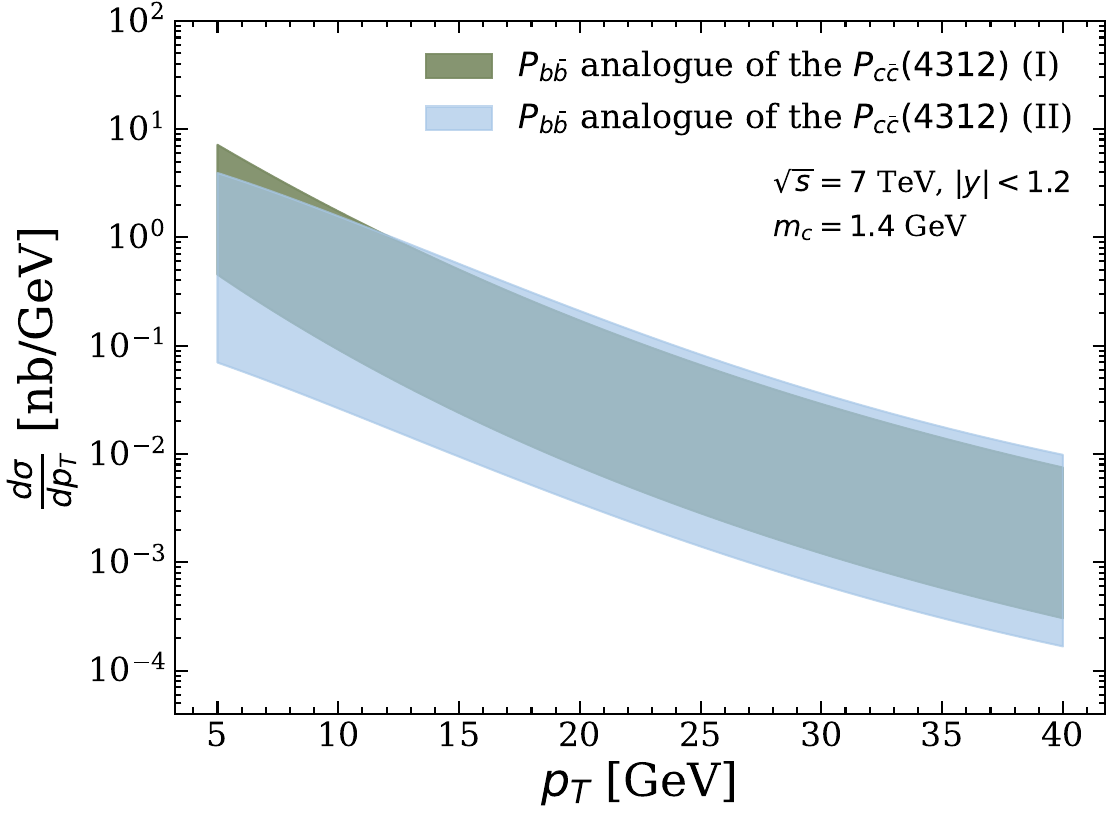}
    \end{subfigure}
    \begin{subfigure}{0.48\textwidth}
        \includegraphics[scale = 0.39]{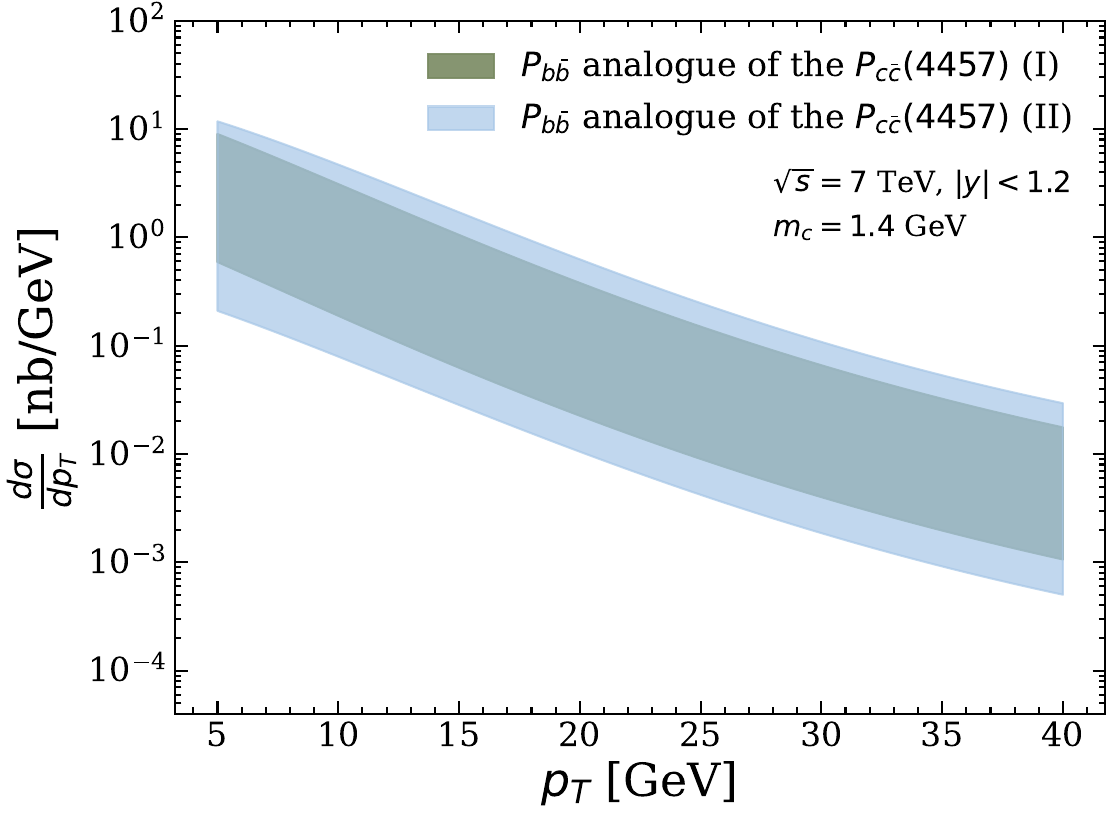}
    \end{subfigure}
    \\
    \begin{subfigure}{0.48\textwidth}
        \includegraphics[scale = 0.39]{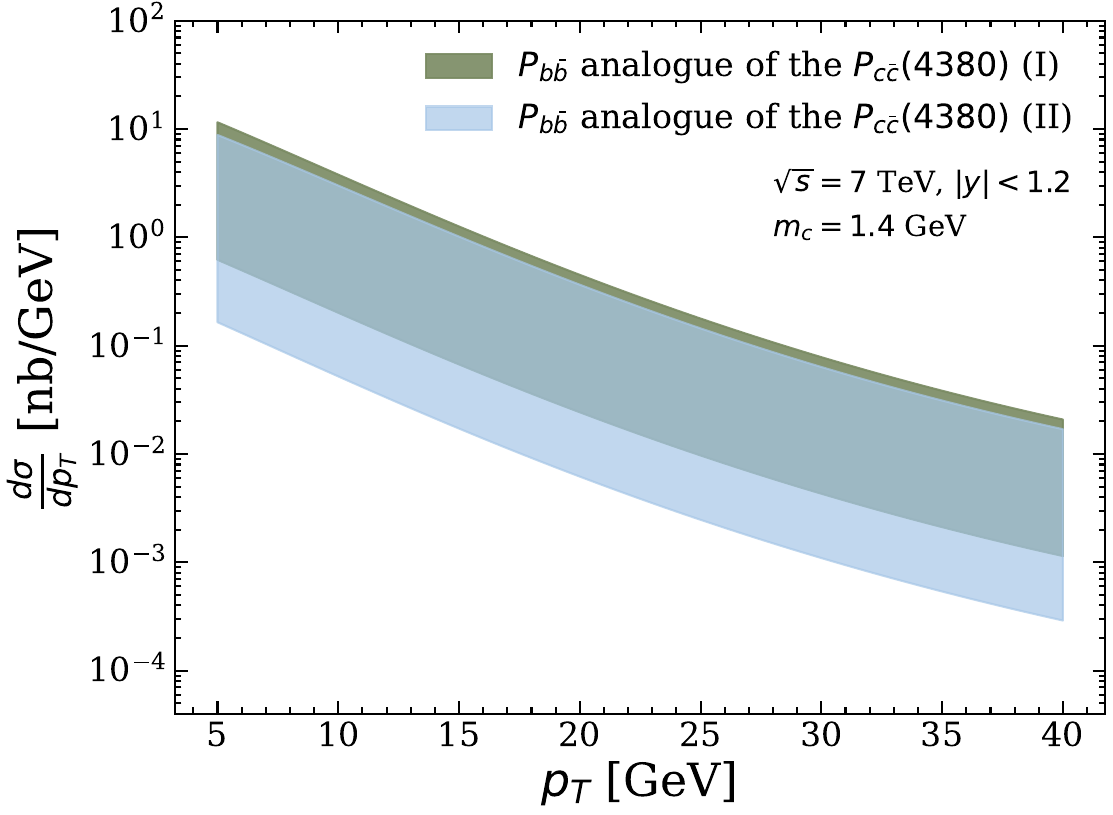}
    \end{subfigure}
    \begin{subfigure}{0.48\textwidth}
        \includegraphics[scale = 0.39]{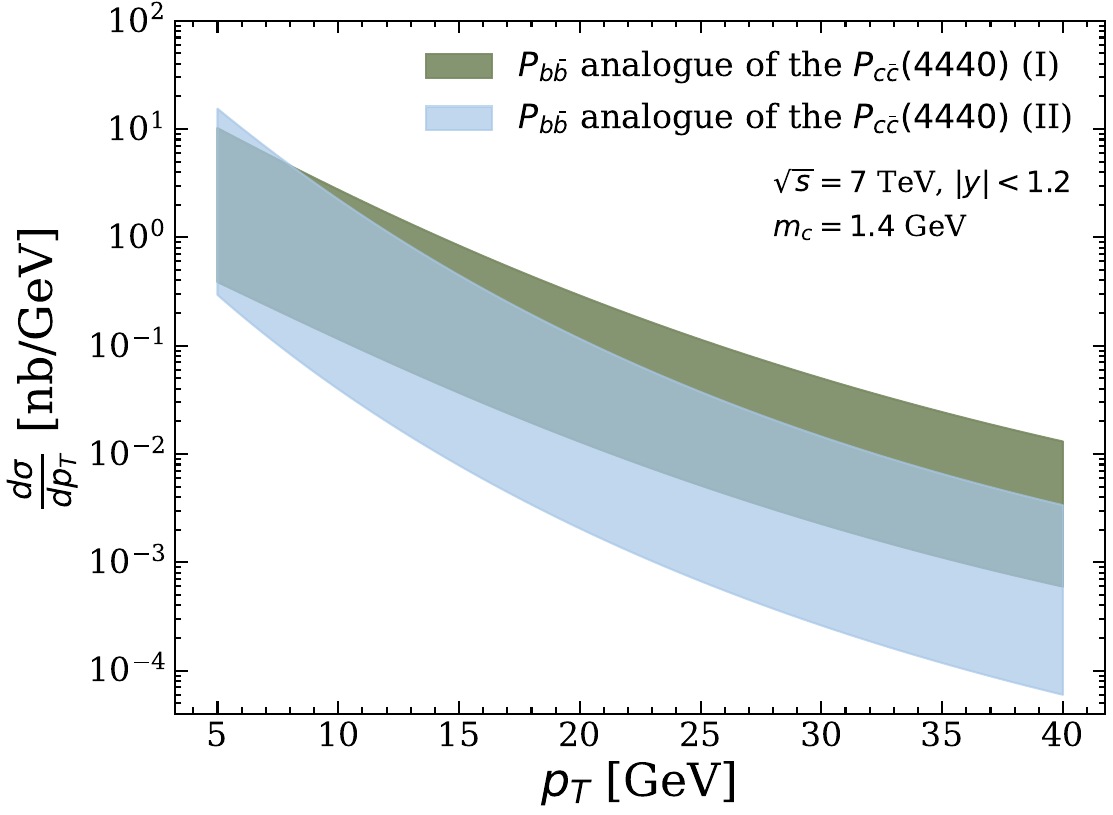}
    \end{subfigure}
    \caption{Predicted prompt inclusive differential hadroproduction cross sections of the bottomonium analogues of the charmonium pentaquark states $P_{c\bar{c}}(4312)^+$,  $P_{c\bar{c}}(4457)^+$,  $P_{c\bar{c}}(4380)^+$ and $P_{c\bar{c}}(4440)^+$ in the scenarios I and II.}
    \label{fig:PentaBottom}
\end{figure}

In the framework of the BOEFT factorization, we can use the matrix elements $\mathcal{M}_{(1/2)_g}$ and $\mathcal{M}_S$ determined in the previous sections on the charmonium pentaquark states to predict the production cross sections of the analogous pentaquark states in the bottomonium sector.
The radial wave functions are computed by solving the Schr\"odinger equations for pentaquarks with the charm mass replaced by the bottom mass~\cite{Brambilla:2025xma}.
We obtain for the wave functions at the origin squared after performing the transformation \eqref{eq:Uphipenta}
\begin{align}
    |\phi_{(1/2)_g}(0)|^2 &\approx  1.14 \times 10^{-1} \, \mathrm{GeV}^3 ,\\
    |\phi_{S}(0)|^2 &\approx 8.02 \times 10^{-2} \, \mathrm{GeV}^3 ,\\
    |\phi_D(0)|^2 &\approx 0,
\end{align}
for scenario I, and
\begin{align}
    |\phi_S(0)|^2 &\approx  9.65 \times 10^{-2} \, \mathrm{GeV}^3 ,\\
    |\phi_D(0)|^2 &\approx 0,
\end{align}
for scenario II.
The differential cross sections for the pentaquarks in the bottomonium sector are displayed in figure~\ref{fig:PentaBottom}.
Uncertainty bands are determined as in the case of the charmonium pentaquarks. 
The green curves show the predictions for scenario I, and the blue curves show those for scenario II.
Both scenarios largely overlap within their error bands.

\section{Conclusions}
\label{sec:conclusions}
In this work, we combine the general framework developed first in~\cite{Brambilla:2020ojz,Brambilla:2021abf} to factorize LDMEs for inclusive production cross sections of quarkonium, then extended to the $\chi_{c1}(3872)$ case in~\cite{Lai:2025tpw}, 
with the Born--Oppenheimer effective field theory~\cite{Berwein:2024ztx} to compute at leading order in the velocity expansion, the inclusive cross sections of the tetraquark state 
$\chi_{c1}(3872)$, the pentaquark states $P_{c\bar c}(4312)^+$, $P_{c\bar c}(4457)^+$, $P_{c\bar c}(4380)^+$ and $P_{c\bar c}(4440)^+$, and their partners in the bottomonium spectrum.
We factorize the LDMEs into the square of the wave functions at the origin and some universal matrix elements that are independent of the heavy quark flavor and mass.
The main factorization formulas are eq. \eqref{octetpNRQCD4} for the $\chi_{c1}(3872)$, and 
eqs. \eqref{eq:LDMEPenta1S08}, \eqref{eq:PentaLDME3S18_final}, \eqref{eq:PentaLDME1S08_4380} and \eqref{eq:PentaLDME3S18_4380} for the pentaquarks in scenario I and eqs. \eqref{eq:LDME3S18PentaScenarioABB}, \eqref{eq:LDME3S18PentaScenarioABBb} and \eqref{eq:LDME1S08PentaScenarioABB}-\eqref{eq:LDME3S18PentaScenarioABBf} for the pentaquarks in scenario II.
We provide an accurate definition of the wave functions as solutions of the BOEFT Schr\"odinger equations describing 
tetraquarks and pentaquarks in the multiplet of interest.
For the $\chi_{c1}(3872)$ and its bottomonium partner, the coupled Schr\"odinger equations are given in eq. \eqref{coupledI0}.
For the considered pentaquarks the relevant Schr\"odinger equations are \eqref{eq:schpenta1} and 
\eqref{eq:schpenta2} in scenario I and only the latter in scenario II.
In all cases, we solve the Schr\"odinger equations and compute numerically the square of the relevant wave functions at the origin.
The nonperturbative input to the Schr\"odinger equations comes from potentials computed in lattice QCD and constrained by symmetries, 
as discussed in detail in~\cite{Brambilla:2024imu,Brambilla:2025xma}.
We also provide the field theoretical definition of the universal matrix elements.

The BOEFT factorization of the LDMEs offers two main advantages. 
First, using a theoretical upper limit and data from $b$-hadron decays, we can constrain the matrix elements in the tetraquark case, eq.~\eqref{eq:NumericalM}, and in the two pentaquark scenarios, eqs.~\eqref{eq:NumericalMIa}, \eqref{eq:NumericalMIb} and \eqref{eq:NumericalMII}.
Combining these constraints with the wave functions at the origin, we determine the LDMEs, which, in turn, we combine with short distance cross sections at next-to-leading order in $\alpha_s$~\cite{Butenschoen:2010rq,Butenschoen:2019lef} to finally obtain hadroproduction cross sections. 
Second, the universality of the matrix elements implies that, once determined, they can be used to make predictions for states of different heavy flavors.

In section~\ref{sec:QQbarqqbar}, we compare the computed cross section for the $\chi_{c1}(3872)$ with experimental data from CMS, ATLAS, and LHCb, 
see figure~\ref{fig:SigmaXc}, and find agreement within the given uncertainties, 
although the predictions lie systematically somewhat above the experimental central values.
Exploiting the universality of the matrix element in the BOEFT factorization of the LDME, 
we then predict the inclusive cross section of the bottom analogue of the $\chi_{c1}(3872)$, 
see figure~\ref{fig:Xb_cross_section}.
In section~\ref{sec:QQbarqqq}, we compute the inclusive cross sections of the pentaquark states $P_{c\bar c}(4312)^+$, $P_{c\bar c}(4457)^+$, $P_{c\bar c}(4380)^+$ and $P_{c\bar c}(4440)^+$, considering the two different scenarios proposed in~\cite{Brambilla:2025xma} and in~\cite{Alasiri:2025roh}.
The predicted pentaquark inclusive cross sections in the two scenarios are shown in the figures~\ref{fig:PentaCharmScI} and~\ref{fig:PentaCharmScII}. 
The predicted inclusive cross sections for the corresponding bottomonium pentaquark states are in figure~\ref{fig:PentaBottom}.
Given the current uncertainties, the two scenarios yield comparable cross sections.

In the last years, the production of the $\chi_{c1}(3872)$ has been the subject of some controversy between those advocating for a compact tetraquark nature of the state~\cite{Bignamini:2009sk} and those supporting the interpretation of the $\chi_{c1}(3872)$ as a loosely bound $D^{*0}\bar{D}^0/D^0\bar{D}^{*0}$ molecule~\cite{Braaten:2018eov}.
It should be emphasized here that the BOEFT description of the $\chi_{c1}(3872)$ does not rely on any a priori assumption on the nature of the $\chi_{c1}(3872)$.
The dynamics of the state is encoded in the potentials entering the coupled Schr\"odinger equations \eqref{coupledI0}. 
The potentials are constrained by symmetries at short and large distances, and by the currently available lattice QCD results~\cite{Bulava:2024jpj}. 
At short distance, the charm-anticharm pair is in a color octet configuration~\cite{Berwein:2024ztx}.
At large distance, the pair opens up in the $D^{*0}\bar{D}^0/D^0\bar{D}^{*0}$ threshold. 
Although the resulting state lies very close to the threshold and has a radius of about 15~fm~\cite{Brambilla:2024imu},
which is consistent with a loosely bound molecule, it is the short-distance interaction that matters for the production mechanism. 
Both short-distance and long-distance interactions are included and properly accounted for in the BOEFT description. 
This is why we end up having a loosely bound state close to the $D^{*0}\bar{D}^0/D^0\bar{D}^{*0}$ threshold that is nevertheless efficiently produced in $pp$ collisions in agreement with LHC measurements, as shown in  figure~\ref{fig:SigmaXc}.

There are several ways in which this work could be improved and expanded.
The computations on the $\chi_{c1}(3872)$ may be improved through various resummations. 
In~\cite{Lai:2025tpw}, both the leading-power resummation and threshold-resummation effects have been considered.
Resummations appear to push the theoretical predictions closer to the experimental central points.
It should be noted, however, that higher-order corrections in the velocity expansion could have a much stronger effect than resummations, at least in the $p_T$ region considered here, if their effect turns out to be larger than the 30\% we have accounted for that follows from having computed the $\chi_{c1}(3872)$ cross section only at leading order in the nonrelativistic expansion.
Going beyond leading order is feasible, but may be of limited phenomenological benefit as it will introduce 
more unknown nonperturbative parameters.
More technically, as pointed out in~\cite{Bodwin:2015iua}, the large leading-power resummation effects originate mostly from partially computed NNLO non-logarithmic contributions; 
NLO plus leading-power resummation contributions alone appear to lead only to small effects. 
Regarding threshold resummation, the expectation is that its effects are small in the relatively low-$p_T$ region considered in this work. 
The pentaquark study requires as most urgent input a lattice QCD determination of the pentaquark potentials.
With this information, we could discriminate between the two scenarios analyzed in the paper and, without assumptions, determine the pentaquark wave functions at the origin. 
As discussed in~\cite{Brambilla:2025xma}, also the determination of the pentaquark $J^P$ quantum numbers may discriminate between the two scenarios.

The study performed here for the $\chi_{c1}(3872)$ can be extended to other XYZ states.
The most straightforward extension appears to be the computation of the inclusive production cross section of quarkonium hybrids.
The lowest lying quarkonium hybrids are supported by the Born--Oppenheimer potential $\Pi_u$, which becomes degenerate with the potential $\Sigma_u^-$ and the gluelump mass of quantum numbers $k^{PC} = 1^{+-}$ at short distance.
At large distance, the $\Sigma_u^-$ potential mixes with an $S$-wave meson $S$-wave antimeson threshold with $k^{PC}=0^{-+}$.
The relevant coupled Schr\"odinger equations can be found in~\cite{Berwein:2024ztx}.
The heavy quark-antiquark pair in a hybrid is in a color octet configuration, hence the production mechanism is,  like in the $\chi_{c1}(3872)$ case, governed at leading order in the velocity expansion by dimension six LDMEs 
of color octet operators either projecting on a spin singlet or a spin triplet state in dependence on the spin of the heavy quark pair in the hybrid. 
The short-distance coefficients entering the NRQCD factorization formula of the cross section are the same as the ones computed for the quarkonium production case, for they involve heavy quark-antiquark pairs in the same color and spin configurations.
Also, inclusive cross sections of tetraquarks of the type $QQ\bar{q}\bar{q}$, most notably the $T_{cc}(3875)^+$~\cite{LHCb:2021vvq}, may be computed in the same framework outlined in this paper.
However, at short distance, the heavy quark pair in a low-lying $QQ\bar{q}\bar{q}$ tetraquark is most likely in a color antitriplet configuation~\cite{Berwein:2024ztx,Brambilla:2024imu}.
This means that the LDMEs contributing at leading order to the inclusive cross section are matrix elements of operators  projecting on a color antitriplet heavy quark pair.
In the case of the $T_{cc}(3875)^+$, the heavy quark pair is also in a spin triplet configuration.
The computation of the cross section requires, therefore, the computation of short-distance coefficients associated with color antitriplet operators, which have no analogue in the quarkonium case.

\section*{Acknowledgments}
S.H. thanks Carlos Louren\c{c}o for useful discussions. 
N.B., S.H., A.M., and A.V. acknowledge support from the DFG cluster of excellence ORIGINS funded by the Deutsche Forschungsgemeinschaft under Germany’s Excellence Strategy-EXC-2094-390783311. 
N.B. acknowledges the Advanced ERC grant ERC-2023-ADG-Project EFT-XYZ.
The work of M.B. is supported by the German Research Foundation DFG through Grant No. BU 3455/1-1 as part of the Research Unit FOR2926. The work of X.-P.~W. is supported by the National Natural
Science Foundation of China under Grant
No.~12135006.

\bibliographystyle{jhep}
\bibliography{references}

@article{Bodwin:1994jh,
    author = "Bodwin, Geoffrey T. and Braaten, Eric and Lepage, G. Peter",
    title = "{Rigorous QCD analysis of inclusive annihilation and production of heavy quarkonium}",
    eprint = "hep-ph/9407339",
    archivePrefix = "arXiv",
    reportNumber = "ANL-HEP-PR-94-24, FERMILAB-PUB-94-073-T, NUHEP-TH-94-5",
    doi = "10.1103/PhysRevD.55.5853",
    journal = "Phys. Rev. D",
    volume = "51",
    pages = "1125--1171",
    year = "1995",
    note = "[Erratum: Phys.Rev.D 55, 5853 (1997)]"
}

@article{Brambilla:2019esw,
    author = "Brambilla, Nora and Eidelman, Simon and Hanhart, Christoph and Nefediev, Alexey and Shen, Cheng-Ping and Thomas, Christopher E. and Vairo, Antonio and Yuan, Chang-Zheng",
    title = "{The $XYZ$ states: experimental and theoretical status and perspectives}",
    eprint = "1907.07583",
    archivePrefix = "arXiv",
    primaryClass = "hep-ex",
    reportNumber = "TUM-EFT 125/19",
    doi = "10.1016/j.physrep.2020.05.001",
    journal = "Phys. Rept.",
    volume = "873",
    pages = "1--154",
    year = "2020"
}

@article{Berwein:2024ztx,
    author = "Berwein, Matthias and Brambilla, Nora and Mohapatra, Abhishek and Vairo, Antonio",
    title = "{Hybrids, tetraquarks, pentaquarks, doubly heavy baryons, and quarkonia in Born--Oppenheimer effective theory}",
    eprint = "2408.04719",
    archivePrefix = "arXiv",
    primaryClass = "hep-ph",
    reportNumber = "TUM-EFT 185/23",
    doi = "10.1103/PhysRevD.110.094040",
    journal = "Phys. Rev. D",
    volume = "110",
    number = "9",
    pages = "094040",
    year = "2024"
}

@article{Brambilla:2024imu,
    author = "Brambilla, Nora and Mohapatra, Abhishek and Scirpa, Tommaso and Vairo, Antonio",
    title = "{Nature of $\chi_{c1}(3872)$ and $T_{cc}^+(3875)$}",
    eprint = "2411.14306",
    archivePrefix = "arXiv",
    primaryClass = "hep-ph",
    reportNumber = "TUM-EFT 193/24",
    doi = "10.1103/pdy7-hvg7",
    journal = "Phys. Rev. Lett.",
    volume = "135",
    number = "13",
    pages = "131902",
    year = "2025"
}

@article{Brambilla:2025xma,
    author = "Brambilla, Nora and Mohapatra, Abhishek and Vairo, Antonio",
    title = "{Unraveling pentaquarks with the Born--Oppenheimer effective theory}",
    eprint = "2508.13050",
    archivePrefix = "arXiv",
    primaryClass = "hep-ph",
    reportNumber = "TUM-EFT 198/25",
    doi = "10.1103/5z3t-rq5f",
    journal = "Phys. Rev. D",
    volume = "112",
    number = "11",
    pages = "114037",
    year = "2025"
}

@article{Brambilla:2020ojz,
    author = "Brambilla, Nora and Chung, Hee Sok and Vairo, Antonio",
    title = "{Inclusive Hadroproduction of $P$-Wave Heavy Quarkonia in Potential Nonrelativistic QCD}",
    eprint = "2007.07613",
    archivePrefix = "arXiv",
    primaryClass = "hep-ph",
    reportNumber = "TUM-EFT 138/20",
    doi = "10.1103/PhysRevLett.126.082003",
    journal = "Phys. Rev. Lett.",
    volume = "126",
    number = "8",
    pages = "082003",
    year = "2021"
}

@article{Brambilla:2021abf,
    author = "Brambilla, Nora and Chung, Hee Sok and Vairo, Antonio",
    title = "{Inclusive production of heavy quarkonia in pNRQCD}",
    eprint = "2106.09417",
    archivePrefix = "arXiv",
    primaryClass = "hep-ph",
    reportNumber = "TUM-EFT 139/20",
    doi = "10.1007/JHEP09(2021)032",
    journal = "JHEP",
    volume = "09",
    pages = "032",
    year = "2021"
}

@article{Brambilla:2022rjd,
    author = "Brambilla, Nora and Chung, Hee Sok and Vairo, Antonio and Wang, Xiang-Peng",
    title = "{Production and polarization of S-wave quarkonia in potential nonrelativistic QCD}",
    eprint = "2203.07778",
    archivePrefix = "arXiv",
    primaryClass = "hep-ph",
    reportNumber = "TUM-EFT 168/22",
    doi = "10.1103/PhysRevD.105.L111503",
    journal = "Phys. Rev. D",
    volume = "105",
    number = "11",
    pages = "L111503",
    year = "2022"
}

@article{Brambilla:2022ayc,
    author = "Brambilla, Nora and Chung, Hee Sok and Vairo, Antonio and Wang, Xiang-Peng",
    title = "{Inclusive production of J/{\ensuremath{\psi}}, {\ensuremath{\psi}}(2S), and {\ensuremath{\Upsilon}} states in pNRQCD}",
    eprint = "2210.17345",
    archivePrefix = "arXiv",
    primaryClass = "hep-ph",
    reportNumber = "TUM-EFT 170/22",
    doi = "10.1007/JHEP03(2023)242",
    journal = "JHEP",
    volume = "03",
    pages = "242",
    year = "2023"
}

@article{Nayak:2005rw,
    author = "Nayak, Gouranga C. and Qiu, Jian-Wei and Sterman, George F.",
    title = "{Fragmentation, factorization and infrared poles in heavy quarkonium production}",
    eprint = "hep-ph/0501235",
    archivePrefix = "arXiv",
    reportNumber = "YITP-SB-05-01",
    doi = "10.1016/j.physletb.2005.03.031",
    journal = "Phys. Lett. B",
    volume = "613",
    pages = "45--51",
    year = "2005"
}

@article{Nayak:2005rt,
    author = "Nayak, Gouranga C. and Qiu, Jian-Wei and Sterman, George F.",
    title = "{Fragmentation, NRQCD and NNLO factorization analysis in heavy quarkonium production}",
    eprint = "hep-ph/0509021",
    archivePrefix = "arXiv",
    reportNumber = "YITP-SB-05-26",
    doi = "10.1103/PhysRevD.72.114012",
    journal = "Phys. Rev. D",
    volume = "72",
    pages = "114012",
    year = "2005"
}

@article{Lai:2025tpw,
    author = "Lai, Wai Kin and Chung, Hee Sok",
    title = "{Hadroproduction data support tetraquark hypothesis for $\chi_{c1}(3872)$}",
    eprint = "2505.06910",
    archivePrefix = "arXiv",
    primaryClass = "hep-ph",
    doi = "10.1103/lkff-d2ph",
    journal = "Phys. Rev. D",
    volume = "112",
    number = "5",
    pages = "054005",
    year = "2025"
}

@article{Beneke:1998ks,
    author = "Beneke, M. and Maltoni, F. and Rothstein, I. Z.",
    title = "{QCD analysis of inclusive B decay into charmonium}",
    eprint = "hep-ph/9808360",
    archivePrefix = "arXiv",
    reportNumber = "CERN-TH-98-240, CMU-9805",
    doi = "10.1103/PhysRevD.59.054003",
    journal = "Phys. Rev. D",
    volume = "59",
    pages = "054003",
    year = "1999"
}

@article{CMS:2013fpt,
    author = "Chatrchyan, Serguei and others",
    collaboration = "CMS",
    title = "{Measurement of the $X$(3872) Production Cross Section Via Decays to $J/\psi \pi^+ \pi^-$ in $pp$ collisions at $\sqrt{s}$ = 7 TeV}",
    eprint = "1302.3968",
    archivePrefix = "arXiv",
    primaryClass = "hep-ex",
    reportNumber = "CMS-BPH-11-011, CERN-PH-EP-2013-014",
    doi = "10.1007/JHEP04(2013)154",
    journal = "JHEP",
    volume = "04",
    pages = "154",
    year = "2013"
}

@article{ATLAS:2016kwu,
    author = "Aaboud, Morad and others",
    collaboration = "ATLAS",
    title = "{Measurements of $\psi(2S)$ and $X(3872) \to J/\psi\pi^+\pi^-$ production in $pp$ collisions at $\sqrt{s} = 8$ TeV with the ATLAS detector}",
    eprint = "1610.09303",
    archivePrefix = "arXiv",
    primaryClass = "hep-ex",
    reportNumber = "CERN-EP-2016-193",
    doi = "10.1007/JHEP01(2017)117",
    journal = "JHEP",
    volume = "01",
    pages = "117",
    year = "2017"
}

@article{Berwein:2015vca,
    author = "Berwein, Matthias and Brambilla, Nora and Tarr{\'u}s Castell{\`a}, Jaume and Vairo, Antonio",
    title = "{Quarkonium Hybrids with Nonrelativistic Effective Field Theories}",
    eprint = "1510.04299",
    archivePrefix = "arXiv",
    primaryClass = "hep-ph",
    reportNumber = "TUM-EFT-45-14",
    doi = "10.1103/PhysRevD.92.114019",
    journal = "Phys. Rev. D",
    volume = "92",
    number = "11",
    pages = "114019",
    year = "2015"
}

@article{Bulava:2024jpj,
    author = "Bulava, John and Knechtli, Francesco and Koch, Vanessa and Morningstar, Colin and Peardon, Michael",
    title = "{The quark-mass dependence of the potential energy between static colour sources in the QCD vacuum with light and strange quarks}",
    eprint = "2403.00754",
    archivePrefix = "arXiv",
    primaryClass = "hep-lat",
    reportNumber = "WUB/24-00",
    doi = "10.1016/j.physletb.2024.138754",
    journal = "Phys. Lett. B",
    volume = "854",
    pages = "138754",
    year = "2024"
}

@article{LHCb:2021ten,
    author = "Aaij, Roel and others",
    collaboration = "LHCb",
    title = "{Measurement of {\ensuremath{\chi}}$_{c1}$(3872) production in proton-proton collisions at $ \sqrt{s} $ = 8 and 13 TeV}",
    eprint = "2109.07360",
    archivePrefix = "arXiv",
    primaryClass = "hep-ex",
    reportNumber = "LHCb-PAPER-2021-026, CERN-EP-2021-182",
    doi = "10.1007/JHEP01(2022)131",
    journal = "JHEP",
    volume = "01",
    pages = "131",
    year = "2022"
}

@article{ParticleDataGroup:2024cfk,
    author = "Navas, S. and others",
    collaboration = "Particle Data Group",
    title = "{Review of particle physics}",
    doi = "10.1103/PhysRevD.110.030001",
    journal = "Phys. Rev. D",
    volume = "110",
    number = "3",
    pages = "030001",
    year = "2024"
}

@article{Alasiri:2025roh,
    author = "Alasiri, Fareed and Braaten, Eric and Bruschini, Roberto",
    title = "{Hidden-heavy pentaquarks and where to find them}",
    eprint = "2507.06991",
    archivePrefix = "arXiv",
    primaryClass = "hep-ph",
    doi = "10.1016/j.physletb.2026.140162",
    journal = "Phys. Lett. B",
    volume = "873",
    pages = "140162",
    year = "2026"
}

@article{LHCb:2015yax,
    author = "Aaij, Roel and others",
    collaboration = "LHCb",
    title = "{Observation of $J/\psi p$ Resonances Consistent with Pentaquark States in $\Lambda_b^0 \to J/\psi K^- p$ Decays}",
    eprint = "1507.03414",
    archivePrefix = "arXiv",
    primaryClass = "hep-ex",
    reportNumber = "CERN-PH-EP-2015-153, LHCB-PAPER-2015-029",
    doi = "10.1103/PhysRevLett.115.072001",
    journal = "Phys. Rev. Lett.",
    volume = "115",
    pages = "072001",
    year = "2015"
}

@article{Belle:2003nnu,
    author = "Choi, S. K. and others",
    collaboration = "Belle",
    title = "{Observation of a narrow charmonium-like state in exclusive $B^\pm \to K^\pm \pi^+ \pi^- J/\psi$ decays}",
    eprint = "hep-ex/0309032",
    archivePrefix = "arXiv",
    doi = "10.1103/PhysRevLett.91.262001",
    journal = "Phys. Rev. Lett.",
    volume = "91",
    pages = "262001",
    year = "2003"
}

@article{LHCb:2013kgk,
    author = "Aaij, R and others",
    collaboration = "LHCb",
    title = "{Determination of the X(3872) meson quantum numbers}",
    eprint = "1302.6269",
    archivePrefix = "arXiv",
    primaryClass = "hep-ex",
    reportNumber = "LHCB-PAPER-2013-001, CERN-PH-EP-2013-017",
    doi = "10.1103/PhysRevLett.110.222001",
    journal = "Phys. Rev. Lett.",
    volume = "110",
    pages = "222001",
    year = "2013"
}

@article{Bodwin:2015iua,
    author = "Bodwin, Geoffrey T. and Chao, Kuang-Ta and Chung, Hee Sok and Kim, U-Rae and Lee, Jungil and Ma, Yan-Qing",
    title = "{Fragmentation contributions to hadroproduction of prompt $J/\psi$, $\chi_{cJ}$, and $\psi(2S)$ states}",
    eprint = "1509.07904",
    archivePrefix = "arXiv",
    primaryClass = "hep-ph",
    doi = "10.1103/PhysRevD.93.034041",
    journal = "Phys. Rev. D",
    volume = "93",
    number = "3",
    pages = "034041",
    year = "2016"
}

@article{LHCb:2015jfc,
    author = "Aaij, Roel and others",
    collaboration = "LHCb",
    title = "{Quantum numbers of the $X(3872)$ state and orbital angular momentum in its $\rho^0 J\psi$ decay}",
    eprint = "1504.06339",
    archivePrefix = "arXiv",
    primaryClass = "hep-ex",
    reportNumber = "LHCB-PAPER-2015-015, CERN-PH-EP-2015-098",
    doi = "10.1103/PhysRevD.92.011102",
    journal = "Phys. Rev. D",
    volume = "92",
    number = "1",
    pages = "011102",
    year = "2015"
}

@article{LHCb:2020fvo,
    author = "Aaij, Roel and others",
    collaboration = "LHCb",
    title = "{Study of the $\psi_2(3823)$ and $\chi_{c1}(3872)$ states in $B^+ \rightarrow \left( J\psi\pi^+\pi^-\right)K^+$ decays}",
    eprint = "2005.13422",
    archivePrefix = "arXiv",
    primaryClass = "hep-ex",
    reportNumber = "CERN-EP-2020-071, LHCb-PAPER-2020-009",
    doi = "10.1007/JHEP08(2020)123",
    journal = "JHEP",
    volume = "08",
    pages = "123",
    year = "2020"
}

@article{Oncala:2017hop,
    author = "Oncala, Rub{\'e}n and Soto, Joan",
    title = "{Heavy Quarkonium Hybrids: Spectrum, Decay and Mixing}",
    eprint = "1702.03900",
    archivePrefix = "arXiv",
    primaryClass = "hep-ph",
    reportNumber = "ICCUB-17-004, NIKHF-2017-005",
    doi = "10.1103/PhysRevD.96.014004",
    journal = "Phys. Rev. D",
    volume = "96",
    number = "1",
    pages = "014004",
    year = "2017"
}

@article{Brambilla:2017uyf,
    author = "Brambilla, Nora and Krein, Gast{\~a}o and Tarr{\'u}s Castell{\`a}, Jaume and Vairo, Antonio",
    title = "{Born--Oppenheimer approximation in an effective field theory language}",
    eprint = "1707.09647",
    archivePrefix = "arXiv",
    primaryClass = "hep-ph",
    reportNumber = "TUM-EFT-69-15",
    doi = "10.1103/PhysRevD.97.016016",
    journal = "Phys. Rev. D",
    volume = "97",
    number = "1",
    pages = "016016",
    year = "2018"
}

@article{Soto:2020xpm,
    author = "Soto, Joan and Tarr{\'u}s Castell{\`a}, Jaume",
    title = "{Nonrelativistic effective field theory for heavy exotic hadrons}",
    eprint = "2005.00552",
    archivePrefix = "arXiv",
    primaryClass = "hep-ph",
    doi = "10.1103/PhysRevD.102.014012",
    journal = "Phys. Rev. D",
    volume = "102",
    number = "1",
    pages = "014012",
    year = "2020",
    note = "[Erratum: Phys.Rev.D 110, 099901 (2024)]"
}

@article{Brambilla:2004jw,
    author = "Brambilla, Nora and Pineda, Antonio and Soto, Joan and Vairo, Antonio",
    title = "{Effective Field Theories for Heavy Quarkonium}",
    eprint = "hep-ph/0410047",
    archivePrefix = "arXiv",
    reportNumber = "IFUM-805-FT, UB-ECM-PF-04-24",
    doi = "10.1103/RevModPhys.77.1423",
    journal = "Rev. Mod. Phys.",
    volume = "77",
    pages = "1423",
    year = "2005"
}

@article{Bignamini:2009sk,
    author = "Bignamini, C. and Grinstein, B. and Piccinini, F. and Polosa, A. D. and Sabelli, C.",
    title = "{Is the X(3872) Production Cross Section at Tevatron Compatible with a Hadron Molecule Interpretation?}",
    eprint = "0906.0882",
    archivePrefix = "arXiv",
    primaryClass = "hep-ph",
    doi = "10.1103/PhysRevLett.103.162001",
    journal = "Phys. Rev. Lett.",
    volume = "103",
    pages = "162001",
    year = "2009"
}

@article{Braaten:2018eov,
    author = "Braaten, Eric and He, Li-Ping and Ingles, Kevin",
    title = "{Estimates of the $X(3872)$ Cross Section at a Hadron Collider}",
    eprint = "1811.08876",
    archivePrefix = "arXiv",
    primaryClass = "hep-ph",
    doi = "10.1103/PhysRevD.100.094024",
    journal = "Phys. Rev. D",
    volume = "100",
    number = "9",
    pages = "094024",
    year = "2019"
}

@article{LHCb:2015oyu,
    author = "Aaij, Roel and others",
    collaboration = "LHCb",
    title = "{First observation of the decay B$_{s}^{0}$  {\textrightarrow} K$_{S}^{0}$ K$^{*}$(892)$^{0}$ at LHCb}",
    eprint = "1506.08634",
    archivePrefix = "arXiv",
    primaryClass = "hep-ex",
    reportNumber = "CERN-PH-EP-2015-144, LHCB-PAPER-2015-018",
    doi = "10.1007/JHEP01(2016)012",
    journal = "JHEP",
    volume = "01",
    pages = "012",
    year = "2016"
}

@article{Brambilla:1999xf,
    author = "Brambilla, Nora and Pineda, Antonio and Soto, Joan and Vairo, Antonio",
    title = "{Potential NRQCD: An Effective theory for heavy quarkonium}",
    eprint = "hep-ph/9907240",
    archivePrefix = "arXiv",
    reportNumber = "CERN-TH-99-199, HEPHY-PUB-716-99, UB-ECM-PF-99-06, UWTHPH-1999-34, UB-ECM-PF-99-13",
    doi = "10.1016/S0550-3213(99)00693-8",
    journal = "Nucl. Phys. B",
    volume = "566",
    pages = "275",
    year = "2000"
}

@article{Brambilla:2000gk,
    author = "Brambilla, Nora and Pineda, Antonio and Soto, Joan and Vairo, Antonio",
    title = "{The QCD potential at $O(1/m)$}",
    eprint = "hep-ph/0002250",
    archivePrefix = "arXiv",
    reportNumber = "CERN-TH-2000-053, UB-ECM-PF-00-03, UWTHPH-1999-47",
    doi = "10.1103/PhysRevD.63.014023",
    journal = "Phys. Rev. D",
    volume = "63",
    pages = "014023",
    year = "2001"
}

@article{Pineda:2000sz,
    author = "Pineda, Antonio and Vairo, Antonio",
    title = "{The QCD potential at $O(1/m^2)$: Complete spin dependent and spin independent result}",
    eprint = "hep-ph/0009145",
    archivePrefix = "arXiv",
    reportNumber = "CERN-TH-2000-197, HD-THEP-00-31",
    doi = "10.1103/PhysRevD.64.039902",
    journal = "Phys. Rev. D",
    volume = "63",
    pages = "054007",
    year = "2001",
    note = "[Erratum: Phys.Rev.D 64, 039902 (2001)]"
}

@article{Brambilla:2002nu,
    author = "Brambilla, Nora and Eiras, Dolors and Pineda, Antonio and Soto, Joan and Vairo, Antonio",
    title = "{Inclusive decays of heavy quarkonium to light particles}",
    eprint = "hep-ph/0208019",
    archivePrefix = "arXiv",
    reportNumber = "CERN-TH-2002-179, IFUM-719-FT, UB-ECM-PF-02-15",
    doi = "10.1103/PhysRevD.67.034018",
    journal = "Phys. Rev. D",
    volume = "67",
    pages = "034018",
    year = "2003"
}

@article{Braaten:2024tbm,
    author = "Braaten, Eric and Bruschini, Roberto",
    title = "{Exotic hidden-heavy hadrons and where to find them}",
    eprint = "2409.08002",
    archivePrefix = "arXiv",
    primaryClass = "hep-ph",
    doi = "10.1016/j.physletb.2025.139386",
    journal = "Phys. Lett. B",
    volume = "863",
    pages = "139386",
    year = "2025"
}

@article{Mateu:2018zym,
    author = "Mateu, Vicent and Ortega, Pablo G. and Entem, David R. and Fern{\'a}ndez, Francisco",
    title = {{Calibrating the Na{\"\i}ve Cornell Model with NRQCD}},
    eprint = "1811.01982",
    archivePrefix = "arXiv",
    primaryClass = "hep-ph",
    reportNumber = "IFT-UAM/CSIC-17-110, IFT-UAM/CSIC-18-109",
    doi = "10.1140/epjc/s10052-019-6808-2",
    journal = "Eur. Phys. J. C",
    volume = "79",
    number = "4",
    pages = "323",
    year = "2019"
}

@article{Bodwin:2007fz,
    author = "Bodwin, Geoffrey T. and Chung, Hee Sok and Kang, Daekyoung and Lee, Jungil and Yu, Chaehyun",
    title = "{Improved determination of color-singlet nonrelativistic QCD matrix elements for S-wave charmonium}",
    eprint = "0710.0994",
    archivePrefix = "arXiv",
    primaryClass = "hep-ph",
    reportNumber = "ANL-HEP-PR-07-48",
    doi = "10.1103/PhysRevD.77.094017",
    journal = "Phys. Rev. D",
    volume = "77",
    pages = "094017",
    year = "2008"
}

@article{LHCb:2021vvq,
    author = "Aaij, Roel and others",
    collaboration = "LHCb",
    title = "{Observation of an exotic narrow doubly charmed tetraquark}",
    eprint = "2109.01038",
    archivePrefix = "arXiv",
    primaryClass = "hep-ex",
    reportNumber = "CERN-EP-2021-165, LHCb-PAPER-2021-031",
    doi = "10.1038/s41567-022-01614-y",
    journal = "Nature Phys.",
    volume = "18",
    number = "7",
    pages = "751--754",
    year = "2022"
}

@misc{hepdata.76839,
    author = "{ATLAS Collaboration}",
    title = "{Measurements of $\psi(2S)$ and $X(3872) \to J/\psi\pi^+\pi^-$ production in $pp$ collisions at $\sqrt{s} = 8$ TeV with the ATLAS detector}",
    howpublished = "{HEPData (collection)}",
    year = 2017,
    note = "\url{https://doi.org/10.17182/hepdata.76839}"
}

@misc{hepdata.60421,
    author = "{CMS Collaboration}",
    title = "{Measurement of the $X$(3872) Production Cross Section Via Decays to $J/\psi \pi^+ \pi^-$ in $pp$ collisions at $\sqrt{s}$ = 7 TeV}",
    howpublished = "{HEPData (collection)}",
    year = 2013,
    note = "\url{https://doi.org/10.17182/hepdata.60421}"
}

@article{Butenschoen:2010rq,
    author = "Butenschoen, Mathias and Kniehl, Bernd A.",
    title = "{Reconciling $J/\psi$ production at HERA, RHIC, Tevatron, and LHC with NRQCD factorization at next-to-leading order}",
    eprint = "1009.5662",
    archivePrefix = "arXiv",
    primaryClass = "hep-ph",
    reportNumber = "DESY-10-101",
    doi = "10.1103/PhysRevLett.106.022003",
    journal = "Phys. Rev. Lett.",
    volume = "106",
    pages = "022003",
    year = "2011"
}

@article{Butenschoen:2019lef,
    author = "Butenschoen, Mathias and Kniehl, Bernd A.",
    title = "{Dipole subtraction at next-to-leading order in nonrelativistic-QCD factorization}",
    eprint = "1909.03698",
    archivePrefix = "arXiv",
    primaryClass = "hep-ph",
    reportNumber = "DESY 19-100, DESY-19-100",
    doi = "10.1016/j.nuclphysb.2019.114843",
    journal = "Nucl. Phys. B",
    volume = "950",
    pages = "114843",
    year = "2020"
}

@article{Brambilla:2001xy,
    author = "Brambilla, Nora and Eiras, Dolors and Pineda, Antonio and Soto, Joan and Vairo, Antonio",
    title = "{New predictions for inclusive heavy quarkonium P wave decays}",
    eprint = "hep-ph/0109130",
    archivePrefix = "arXiv",
    reportNumber = "CERN-TH-2001-249, IFUM-692-FT, TTP-01-20, UB-ECM-PF-01-10",
    doi = "10.1103/PhysRevLett.88.012003",
    journal = "Phys. Rev. Lett.",
    volume = "88",
    pages = "012003",
    year = "2002"
}

@article{Butenschoen:2013pxa,
    author = "Butenschoen, Mathias and He, Zhi-Guo and Kniehl, Bernd A.",
    title = "{NLO NRQCD disfavors the interpretation of X(3872) as $\chi_{c1}(2P)$}",
    eprint = "1303.6524",
    archivePrefix = "arXiv",
    primaryClass = "hep-ph",
    reportNumber = "DESY-13-057, UWTHPH-2013-6",
    doi = "10.1103/PhysRevD.88.011501",
    journal = "Phys. Rev. D",
    volume = "88",
    pages = "011501",
    year = "2013"
}

@article{Meng:2013gga,
    author = "Meng, Ce and Han, Hao and Chao, Kuang-Ta",
    title = "{X(3872) and its production at hadron colliders}",
    eprint = "1304.6710",
    archivePrefix = "arXiv",
    primaryClass = "hep-ph",
    doi = "10.1103/PhysRevD.96.074014",
    journal = "Phys. Rev. D",
    volume = "96",
    number = "7",
    pages = "074014",
    year = "2017"
}

@article{Butenschoen:2019npa,
    author = "Butenschoen, Mathias and He, Zhi-Guo and Kniehl, Bernd A.",
    title = "{Deciphering the $X(3872)$ via its polarization in prompt production at the CERN LHC}",
    eprint = "1906.08553",
    archivePrefix = "arXiv",
    primaryClass = "hep-ph",
    reportNumber = "DESY-19-097",
    doi = "10.1103/PhysRevLett.123.032001",
    journal = "Phys. Rev. Lett.",
    volume = "123",
    number = "3",
    pages = "032001",
    year = "2019"
}

@article{Pumplin:2002vw,
    author = "Pumplin, J. and Stump, D. R. and Huston, J. and Lai, H. L. and Nadolsky, Pavel M. and Tung, W. K.",
    title = "{New generation of parton distributions with uncertainties from global QCD analysis}",
    eprint = "hep-ph/0201195",
    archivePrefix = "arXiv",
    reportNumber = "MSU-HEP-011101",
    doi = "10.1088/1126-6708/2002/07/012",
    journal = "JHEP",
    volume = "07",
    pages = "012",
    year = "2002"
}

@InCollection{Crowder2020,
  author    = {Crowder, Stephen and Delker, Collin and Forrest, Eric and Martin, Nevin},
  title     = {Monte Carlo Methods for the Propagation of Uncertainties},
  booktitle = {Introduction to Statistics in Metrology},
  publisher = {Springer International Publishing},
  address   = {Cham},
  year      = {2020},
  pages     = {153--180},
  doi       = {10.1007/978-3-030-53329-8_8},
  isbn      = {978-3-030-53329-8}
}

@article{Lafferty:1994cj,
    author = "Lafferty, G. D. and Wyatt, T. R.",
    title = "{Where to stick your data points: The treatment of measurements within wide bins}",
    reportNumber = "MAN-HEP-94-3, CERN-PPE-94-72, CERN-PPE-94-072",
    doi = "10.1016/0168-9002(94)01112-5",
    journal = "Nucl. Instrum. Meth. A",
    volume = "355",
    pages = "541--547",
    year = "1995"
}

\end{document}